\documentclass[12pt]{article}

\usepackage{graphicx}
\usepackage{epsfig}
\usepackage{multirow}

\usepackage{amsmath}
\usepackage{amssymb}
\usepackage{ntheorem}
\usepackage[ansinew]{inputenc}
\usepackage{graphicx}
\usepackage{epsfig}
\usepackage{dsfont}
\usepackage{setspace}

\usepackage{color}

\usepackage[authoryear]{natbib}
\usepackage{latexsym}
\newcommand{\NND}{\mbox{NND}}
\newcommand{\la}{\lambda}
\newcommand{\ga}{\gamma}

\newcommand{\vp}{\varphi}
\newcommand{\supp}{\mbox{supp}}

\newcommand{\R}{\mathbb{R}}
\newcommand{\bea}{\begin{eqnarray}}
\newcommand{\eea}{\end{eqnarray}}
\newcommand{\be}{\begin{equation}}
\newcommand{\ee}{\end{equation}}
\newcommand{\beo}{\begin{equation*}}
\newcommand{\eeo}{\end{equation*}}
\newcommand{\beao}{\begin{eqnarray*}}
\newcommand{\eeao}{\end{eqnarray*}}
\newcommand{\ba}{\begin{array}}
\newcommand{\ea}{\end{array}}

\newtheorem{satz}{Theorem}[section]

\newtheorem{defin}[satz]{Definition}
\newtheorem{kor}[satz]{Corollary}
\newtheorem{exam}[satz]{Example}
\newtheorem{rem}[satz]{Remark}

\begin{document}

\title{Optimal designs for comparing curves}

\author{
 {\small Holger Dette, Kirsten Schorning} \\
{\small Ruhr-Universit\"at Bochum} \\
{\small Fakult\"at f\"ur Mathematik} \\
{\small 44780 Bochum, Germany} \\
{\small e-mail: holger.dette@rub.de, kirsten.schorning@rub.de}\\
% {\small FAX: +49 234 3214 559}\\
}

\maketitle

\begin{abstract}
We consider the optimal design problem for a comparison of two regression curves, which is used to establish the similarity between the
dose response relationships of  two groups.  An optimal pair of designs minimizes the width of the confidence band for the difference between the two regression functions. Optimal design theory (equivalence theorems, efficiency bounds) is developed for this non standard design problem and for some commonly used dose response models optimal designs are found explicitly. The results are illustrated in several examples modeling dose response relationships. It is demonstrated that the optimal pair of designs for the comparison of the regression curves is {\bf not} the pair of the optimal designs for the individual models.   In particular it is shown that the use of the optimal designs proposed in this paper instead of commonly used "non-optimal" designs yields a reduction of the width of the confidence band by more than $50\%$.
\end{abstract}

\bigskip
%\noindent { Keywords and Phrases: locally optimal design; admissible design; Chebyshev system; principle representations; moment spaces; complete classes of designs}


\noindent \section{Introduction}
\def\theequation{1.\arabic{equation}}
\setcounter{equation}{0}
\label{sec1}

\begin{figure}
 \begin{center}
 \includegraphics[width=0.4\textwidth]{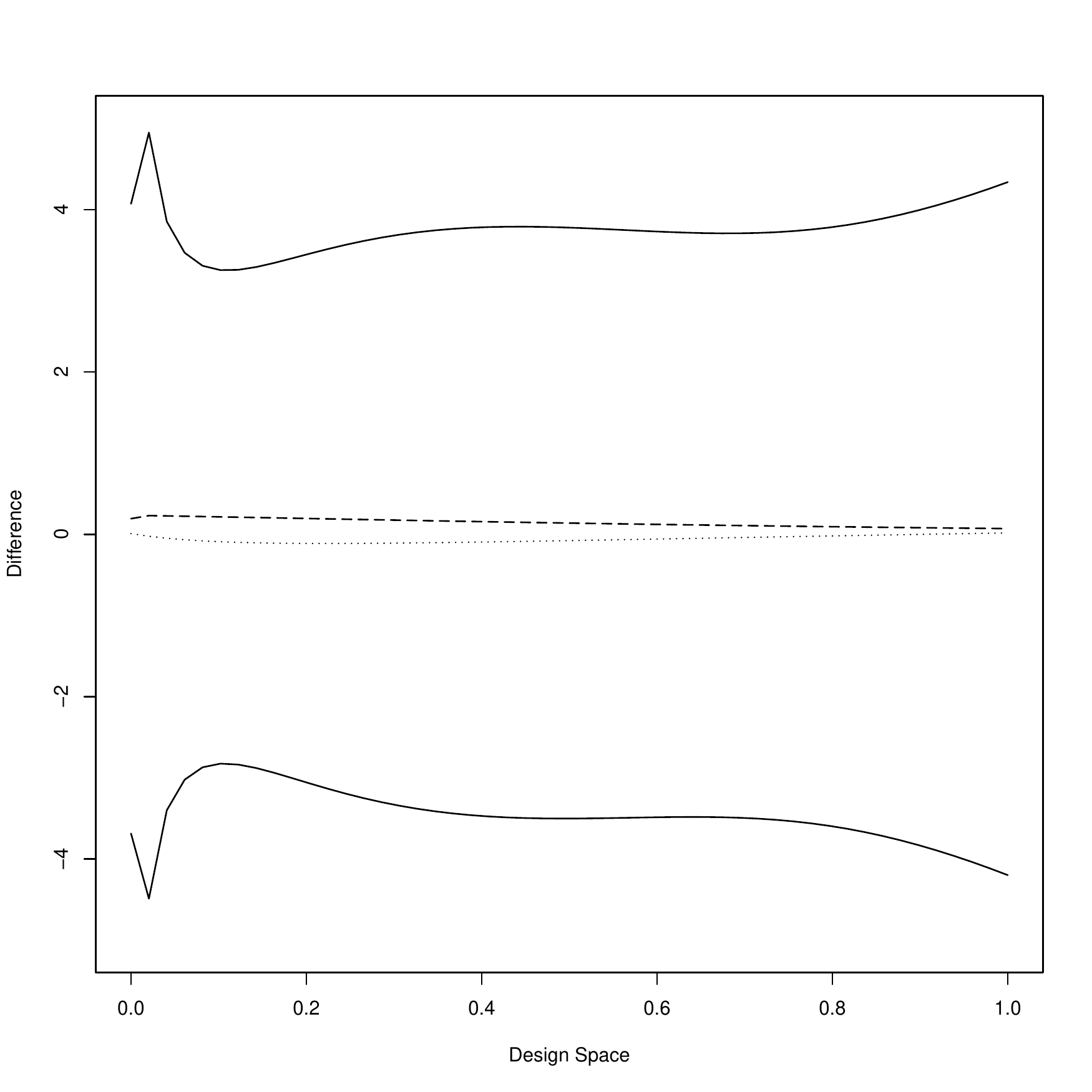} ~~
  \includegraphics[width=0.4\textwidth]{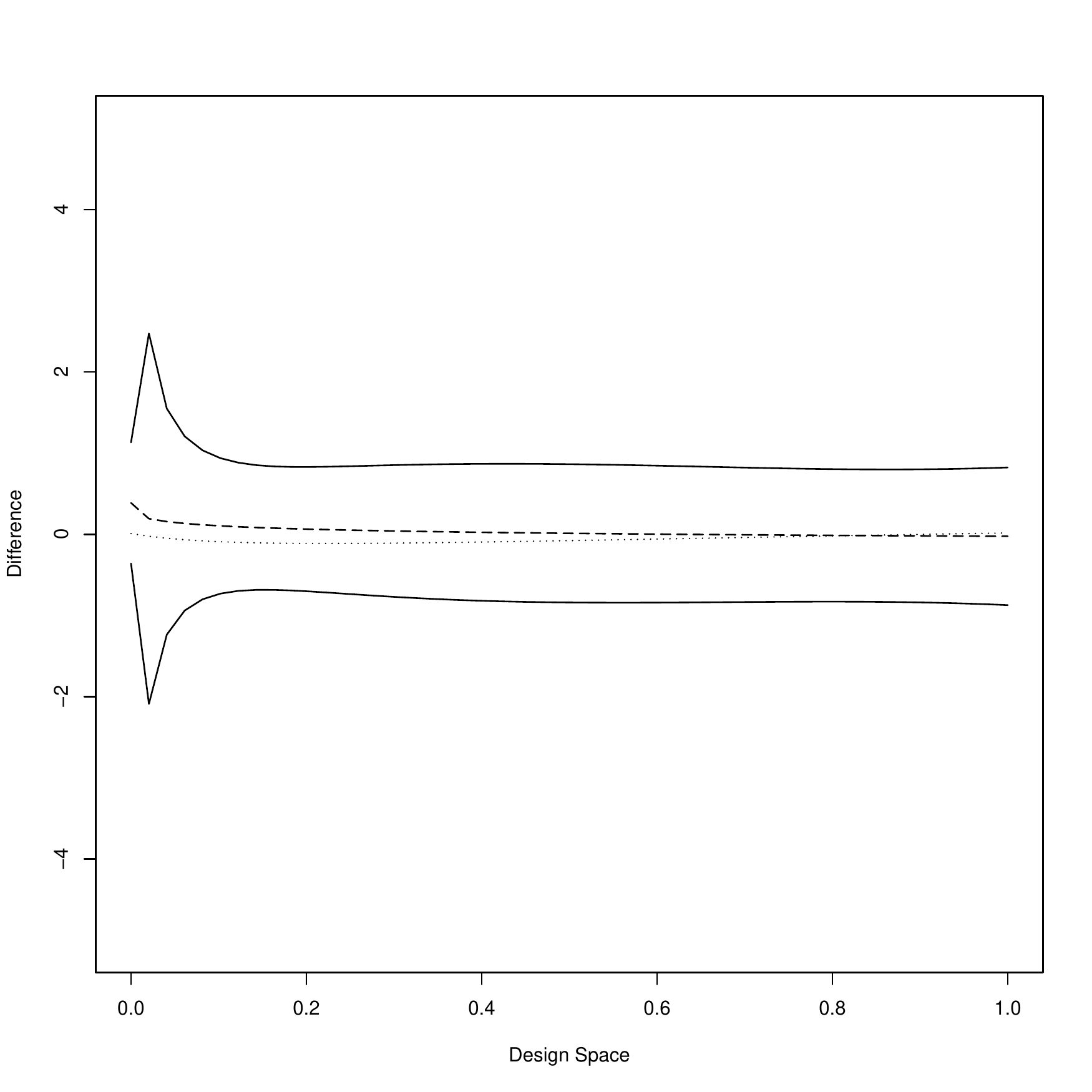} 
  \caption{\it Confidence bands for the difference of the EMAX and loglinear model using a standard design (left panel) and the optimal design (right panel). \label{fig:kb_intro}}
  \end{center}
 \end{figure}

An important problem in many scientific research areas is the comparison of two regression models that describe the relation between a common response and the same covariates for two groups. Such comparisons are typically used to establish the non-superiority of one model to the other or to check whether the difference between two regression models  can be neglected. These investigations have important applications in drug development and several methods for assessing non-superiority, non-inferiority or equivalence have been proposed in the recent literature [for recent references
 see for example \cite{liubrehaywynn2009}, \cite{gsteiger2011}, \cite{liujamzhang2011} among  others]. 
For example, if the ``equivalence'' between two regression models describing the 
dose response relationships in the groups individually has been established
subsequent inference
 in drug development
  could be based on the combined samples. This results in more precise estimates of the relevant parameters, for example the minimum effective dose (MED)  or the median effective dose (ED50).
A common approach in all these references is to estimate regression curves in the different samples and to investigate the maximum or an $L_2$-distance  (taking over the possible range of the covariates) of the difference between these estimates (after an appropriate standardization by a variance estimate). Comparison of curves problems  have been investigated in linear and nonlinear models [see  \cite{liubrehaywynn2009}, \cite{gsteiger2011}, \cite{liujamzhang2011}] and also in nonparametric regression models [see for example  \cite{halhar1990}, \cite{harmar1990} and \cite{detneu2001}]. \\
This paper is devoted to the construction of efficient designs for the comparison of two parametric curves. Although the consideration of optimal designs for dose response models has found considerable interest in the recent literature [see for example \cite{fedleo2001}, \cite{laueretal1997}, \cite{wang2006}, \cite{zhuwong2000}, \cite{debrpepi2008}, \cite{dragetal2010} and \cite{detborbre2013} among many  others], we are not aware of any work on   design of experiments for the comparison of two parametric regression curves. 
However, the effective planning of the experiments in the comparison of curves  will yield to a substantially more accurate statistical inference. We demonstrate these advantages here in a 
 small example to motivate the theoretical investigations of the following sections.
More examples illustrating the advantages of optimal design theory in the context of comparing curves can be found in Section \ref{sec5}. \\
  \cite{gsteiger2011} proposed  a    confidence band for the difference of two regression curves, say
$m_1( \cdot, \vartheta_1) - m_2 ( \cdot , \vartheta_2) $, using a bootstrap approach,  where $m_1 (\cdot,  \vartheta_1)  $ and   $m_2 (\cdot,  \vartheta_2)  $  are two parametric regression 
models with parameters  $ \vartheta_1$ and $\vartheta_2$, respectively.
This band is then used to decide at a controlled type I error for the similarity of the curves, that is for a test of the hypotheses 
\begin{equation}\label{similar}
H_0: ~\sup_{t \in {\cal Z}} | m_1 (t, \vartheta_1) - m_2 (t, \vartheta_2) | > \Delta ~~~\mbox{versus} ~~~H_1: ~\sup_{t \in {\cal Z}} | m_1 (t, \vartheta_1) - m_2 (t, \vartheta_2) | \le \Delta ~,
\end{equation}
where $ {\cal Z}$ is a region of interest for the predictor (for example the dose range in a dose finding study) and $\Delta >0$ a pre-specified constant, for which the difference 
between the two models is considered as negligible. Roughly speaking these authors considered the curves as similar if the maximum (minimum) of the upper (lower) confidence bound is smaller (larger) 
than $\Delta$ ($-\Delta$). 
In Figure \ref{fig:kb_intro}   we  display uniform confidence bands for the difference of an  EMAX and a loglinear model, which were investigated by  \cite{bretz2005} for modeling dose response relationships. The sample sizes for both groups are $n_1=100$ and $n_2=100$, respectively.
The left hand part of Figure \ref{fig:kb_intro} shows
 the average of uniform confidence bands (solid lines), the average estimate of the difference calculated by  $100$ simulation runs (dashed line)
 and the ''true'' difference of the two functions (dotted  line), where patients are allocated to the different dose levels according to a standard design  (for details see Section \ref{sec5}). The corresponding confidence bands
calculated from observations sampled with  respect to the  optimal designs derived in this paper are shown in the right part of Figure \ref{fig:kb_intro} and
we observe that an optimal design yields to  substantially narrower confidence bands for the difference of the regression functions. 
As a consequence tests of the hypotheses  of the form \eqref{similar} are substantially more powerful. In other words: we actually decide more often for the similarity of the curves, resulting in a more accurate statistical inference by finally merging the information of the two groups.  \\
The  present paper is motivated by  observations of this type and will address the problem of constructing optimal designs of  experiments  for the comparison of curves.   
 Some terminology (for the comparison of two parametric curves) will be introduced in Section \ref{sec2}, where we also give an introduction to optimal design theory in the present context. The particular difference to the classical setup is that for the comparison of two curves two designs have to be chosen 
 simultaneously (each for one group or regression model). A pair of optimal designs minimizes an integral or the maximum of the variance of the prediction for the difference of the two regression curves calculated in the common region of interest.  \\
 Section \ref{sec3} is devoted to some optimal design theory and we derive particular equivalence theorems corresponding to the new optimality criteria
 and a lower bound for the efficiencies, which can be used without knowing the optimal designs. 
 It turns out that in general the optimal pair of designs is not the pair of the optimal designs in the individual models.
 We also consider the problem where a design (for one curve) is fixed and only the design for estimating the second curve has to be determined, such that a most efficient comparison of the curves can be conducted.  \\
In general, the problem of constructing optimal designs is very difficult and has to be solved numerically in most cases of practical interest. Some analytical results are given in Section \ref{sec4}, where we deal with the problem of extrapolation. We first derive an explicit solution for weighted polynomial regression models of arbitrary degree, which is of its own interest. These results
 are then used to  determine optimal designs for comparing curves modeled by the commonly used Michaelis Menten, EMAX and loglinear model.
In  Section \ref{sec5} we use the developed theory to 
 investigate specific optimal design problems for the comparison of nonlinear regression models, which are frequently used in drug development.
 In particular we demonstrate by means of a  simulation study that the derived optimal designs yield substantially narrower confidence bands (and as a consequence
 more powerful tests for the hypotheses \eqref{similar}). Finally, in Section \ref{sec6a} we briefly indicate how the results can be generalized if
 optimization can also be performed with respect to the allocation of patients to the different groups, while
 all proofs  are deferred to an Appendix in Section \ref{sec6}.  \\
  For the sake of brevity we restricted ourselves to locally optimal designs
which require a-priori information about the unknown model parameters if the models are nonlinear 
[see \cite{chernoff1953}]. In several situations   preliminary knowledge regarding the unknown parameters of a nonlinear model is available, and 
the application of
locally optimal designs is well justified.  A typical example are phase II clinical dose finding trials, where some useful knowledge is
already available from phase I [see \cite{debrpepi2008}]. Moreover, these designs can
be used  as benchmarks for commonly used designs, and locally optimal designs serve as basis for constructing optimal 
designs with respect to more sophisticated optimality
criteria,  which
are robust against a misspecification of the unknown parameters [see \cite{pronwalt1985}  or \cite{chaver1995}, \cite{dette1997} among others]. 
Following this line of research the methodology introduced in the present paper can be further developed to address uncertainty in the preliminary information for the unknown parameters. 

 \noindent \section{Comparing parametric curves} \label{sec2}
\def\theequation{2.\arabic{equation}}
\setcounter{equation}{0}

Consider the regression  model
\begin{equation} 
Y_{ijk}=m_i(t_{ij}, \vartheta_i) + \varepsilon_{ijk} \ ; \quad i=1,2; \quad j=1,\dots,\ell_i \ ; \quad k=1,\dots,n_{ij},
\label{mod}
\end{equation}
where $\varepsilon_{ijk}$ are independent random variables, such that $\varepsilon_{ijk} \sim \mathcal{N}(0,\sigma^2_i)$, $ i=1,2$. This means that two groups $(i=1,2)$  are investigated and in each group observations are taken at $\ell_i$ different experimental conditions $t_{i1},\dots,t_{i,\ell_i}$, which vary in the design space (for example the  dose range) $ \mathcal{X}\subset \mathbb{R}$, and
$n_{ij}$ observations are taken at each $t_{ij} \ (i=1,2; \ j=1,\dots,\ell_i)$. Let $n_i = \sum^{\ell_i}_{j=1} n_{ij}$ denote the total number of observations in
group $i \ (=1,2)$ and $n=n_1+n_2$ the total sample size.  Two regression models $m_1$ and $m_2$ with $d_1$- and $d_2$-dimensional parameters
$\vartheta_1$ and $\vartheta_2$ are used  to describe the dependency between response and predictor in the two groups.
For asymptotic arguments we assume that $\lim_{n_i \to \infty} \frac {n_{ij}}{n_i}=\xi_{ij} \in (0,1)$ and collect this information 
 in the matrix
$$
\xi_i = \left (
\begin{array}{ccc}
t_{i1} & \dots & t_{i \ell_i} \\
\xi_{i1} & \dots & \xi_{i \ell_i}
\end{array}
\right), \quad i=1, 2.
$$
Following \cite{kiefer1974} we call $\xi_i$  an approximate design on the design space $\mathcal{X}$. This means that the support points ${t_{ij}}$  define the distinct experimental conditions  where observations are to be taken and the weights $\xi_{ij}$ represent the relative proportion of observations at the corresponding support point   $t_{ij}$ (in each group).   If an approximate design is given and  $n_i$ observations can be taken, a rounding procedure is applied
to obtain  integers $n_{ij} $ ($i=1,2$, $j=1,\ldots,{\ell_i})$   from the not necessarily integer valued quantities
$\xi_{ij}n_i$  [see \cite{pukrie1992}].   \\
If  observations are taken according to an approximate design and an appropriate rounding procedure has been applied such that
$\lim_{n_i \to \infty} \frac {n_{ij}}{n_i}=\xi_{ij} \in (0,1)$,
then under the common assumptions of regularity, the maximum likelihood estimates $\hat \vartheta_1, \hat \vartheta_2$ of both samples satisfy
$$
\sqrt{n_i} (\hat \vartheta_i - \vartheta_i)  \stackrel{\mathcal{D}} \longrightarrow \mathcal{N}(0, \sigma^2_i M^{-1}_i (\xi_i, \vartheta_i)), \quad i=1,2~,
$$
where the symbol $  \stackrel{\mathcal{D}}\longrightarrow  $ denotes  weak convergence,
$$
M_i (\xi_i, \vartheta_i) = \int_{\mathcal{X}} f_i (t) f^T_i (t) d \xi_i (t) 
$$
is the information matrix of the design $\xi_i$ in model $m_i$ and $f_i(t)= \frac {\partial}{\partial \vartheta_i}m_i (t,\vartheta_i) \in \mathbb{R}^{d_i}  $ is the gradient of $m_i$ 
with respect to the parameter $\vartheta_i \in \mathbb{R}^{d_i} \ (i=1,2)$. Note that under different distributional assumptions on the errors $\varepsilon_{ijk}$ in model \eqref{mod} similar
statements can be derived with different covariance matrices in the asymptotic distribution. \\
By the delta method we obtain for the difference of the prediction $m_1 (t, \hat \vartheta_1) -m_2 (t, \hat \vartheta_2)$ at the point $t$
\beo
\sqrt{n} \bigl(m_1(t,\hat \vartheta_1) - m_2 (t, \hat \vartheta_2) - (m_1(t,\vartheta_1)-m_2(t,\vartheta_2) \bigr) \stackrel{\mathcal{D}} \longrightarrow \mathcal{N} \bigl(0, \vp(t, \xi_1, \xi_2) \bigr),
\eeo
where the function $\vp$ is defined by 
\be \label{phi_def}
 \vp(t, \xi_1, \xi_2) 
 %=\vp_1(t, \xi_1) + \vp_2(t, \xi_2) 
 =\frac{\sigma^2_1}{\ga_1} f^T_1 (t) M^{-1}_1 (\xi_1, \vartheta_1) f_1(t) + \frac{\sigma^2_2}{\ga_2}f^T_2 (t) M^{-1}_2 (\xi_2, \vartheta_2) f_2(t).
\ee
For these calculations we assume in particular the existence
$$
\gamma_i = \lim_{n \to \infty} \frac {n_i}{n} \in (0,1), \qquad i=1, 2
$$
with $\ga_1 + \ga_2=1$ and that $m_1, \, m_2$ are continuously differentiable with respect to 
the parameters $\vartheta_1, \vartheta_2$. Therefore  the asymptotic variance of the prediction $m_1(t,\hat \vartheta_1)- \hat m_2(t,\hat \vartheta_2)$  at an experimental condition $t$  
 is given by $ \vp(t, \xi_1, \xi_2) $, where  $\xi= (\xi_1, \xi_2)$ is the pair of designs under consideration.
\cite{gsteiger2011} used this result to obtain a simultaneous confidence band for the difference of the two curves. More precisely, if $\mathcal{Z}$ is a range where the two curves should be compared (note that in contrast to \cite{gsteiger2011} here the set $\mathcal{Z}$ does not necessarily coincide with the design space $\mathcal{X}$) the confidence band is defined by
\begin{equation}\label{confset}
 \hat T \equiv \sup_{t \in \mathcal{Z}} \ \frac {|m_1(t,\hat \vartheta_1)-m_2(t,\hat \vartheta_2) - (m_1(t,\vartheta_1)-m_2(t,\vartheta_2))|}{\{ \frac {  \hat \sigma^2_1}{\gamma_1} \hat f_1 (t) M^{-1}_1 (\xi_1,\hat \vartheta_1)\hat f_1(t) + \frac {\hat \sigma^2_2}{\gamma_2}\hat f_2 (t) M^{-1}_2 (\xi_2,\hat \vartheta_2) \hat f_2 (t) \}^{1/2} } \leq D.
\end{equation}
Here, $\hat \sigma^2_1, \hat \sigma^2_2, \hat f_1, \hat f_2$ denote estimates of the quantities $\sigma^2_1, \sigma^2_2, f_1, f_2$, respectively and  the constant $d$ is chosen, such that $\mathbb{P}(\hat T \leq D) \approx 1-\alpha$. Note that \cite{gsteiger2011} proposed the parametric bootstrap for this purpose. Consequently,
 a ``good'' design, more precisely, a pair   $\xi=(\xi_1, \xi_2)$ of two designs on $\mathcal{X}$, should make  the width of this band as small as possible at each $t \in \mathcal{Z}$. This corresponds to a simultaneous minimization of the asymptotic variance in \eqref{phi_def} with respect to the choice of  the designs $\xi_1$ and $\xi_2$. 
  Obviously, this is only possible in rare circumstances and we propose to minimize a norm of the function $\vp$ as a design criterion.
For a precise definition of the optimality criterion we assume
 that the  set  $ \mathcal{Z} $ contains at least $d=\max \{d_1,d_2\}$ points, say $t_1, \ldots ,t_d$, such that the vectors $f_1(t_1),\ldots , f_1(t_{d_1})$ 
 and  $f_2(t_1),\ldots , f_2(t_{d_2})$ are linearly independent in $\R^{d_1}$  and  $\R^{d_2}$, respectively. It then follows, that a pair of designs $\xi=(\xi_1,\xi_2)$, which allows 
 to predict the regression $m_1$  at the points $t_1,\ldots , t_{d_1}$  and $m_2$ at  $t_1,\ldots , t_{d_2}$, must have nonsingular information matrices 
 $M_1 (\xi_1, \vartheta_1)$ and   $M_2 (\xi_2, \vartheta_2)$,  respectively.
 % [see also  \cite{dettobri1999}]. 
 This means that optimization will be restricted to the class
 of all designs $\xi_1$ and $\xi_2$  with non-singular information matrices throughout this paper.\\
   A worst case  criterion is to minimize
\bea \label{infty_crit}
 \mu_{\infty} (\xi) & =&  \mu_{\infty} (\xi_1, \xi_2)  = \sup_{t \in \mathcal{Z}} \{ \vp(t, \xi_1, \xi_2)\}  \\
 &=&
\sup_{t\in \mathcal{Z}}  \Bigl( \frac  {\sigma^2_1}{\ga_1} f^T_1(t) M^{-1}_1 (\xi_1, \vartheta_1) f_1(t) +
\frac  {\sigma^2_2}{\ga_2} f^T_2(t) M^{-1}_2 (\xi_2, \vartheta_2) f_2(t)  \Bigr) \nonumber 
\eea
with respect to $\xi=(\xi_1, \xi_2) $ over a region of interest $\mathcal{Z}$. 
Alternatively, one could
use an $L_p$-norm 
\be \label{p_crit}
 \mu_{p} (\xi)= \mu_p (\xi_1, \xi_2)= \Bigl( \int_{\mathcal{Z}} \vp^p(t, \xi_1, \xi_2)  d \lambda(t)\Bigr)^{1/p}
\ee
of the function $\varphi$ defined in \eqref{phi_def}   with respect to a given measure $\la$  on the region $\mathcal{Z}$ ($p\in [1, \infty)$), 
where the measure $\lambda $ has at least  $d=\max \{d_1,d_2\}$ support points, say $t_1, \ldots ,t_d$, such that the vectors $f_1(t_1),\ldots , f_1(t_{d_1})$ 
 and  $f_2(t_2),\ldots , f_2(t_{d_2})$ are linearly independent in $\R^{d_1}$  and  $\R^{d_2}$, respectively.
 
%For this purpose the support of $\la$ is assumed to be larger than $\max\{s_1, s_2 \} $.%which ensures the admissibility of the optimal design tuple.
%\\
%An optimal design tuple $\xi=(\xi_1, \xi_2)$ is defined as follows.
%  Definition of mu_p
\begin{defin}\label{def_mu}
For $p\in [1, \infty]$ a pair of designs  $\xi^{\star, p} = (\xi^{\star, p}_1, \xi^{\star, p}_2)$ is called locally $\mu_p$-optimal  design (for the comparison of the curves
$m_1$ and $m_2$) if it minimizes the function $
\mu_p (\xi_1, \xi_2)$ over the space of all approximate pairs of designs $(\xi_1, \xi_2)$ on $\mathcal{X} \times \mathcal{X}$  with nonsingular information matrices $M(\xi_1, \vartheta_1)$,  $M(\xi_2, \vartheta_2)$.
\end{defin}

\begin{rem} ~~\\  \rm 
(a) The space $\mathcal{Z}$ does not necessarily coincide  with the design space $\mathcal{X}$. The  special case 
$\mathcal{Z} \cap  \mathcal{X} = \emptyset$ corresponds to  the problem of extrapolation and  will be discussed in more detail in Section \ref{sec4}. 

 \smallskip
(b)  If one requires $\xi_1 = \xi_2$ (for example by logistic reasons) and $\mathcal{Z}=\mathcal{X}$ the criterion $\mu_\infty$    is equivalent to
the  weighted  $D$-optimality criterion $ (\det  M_1(\xi , \vartheta_1) )^{\omega_1} (\det  M_2(\xi , \vartheta_2) )^{\omega_2}, $ where the weights are given by 
$ \omega_1 = \frac {\sigma^2_1}{\ga_1} $ and $\omega_2 = \frac {\sigma^2_2}{\ga_2}.$ Criteria of this type have been
studied intensively in the literature [see \cite{laustu1985}, \cite{dette1990}, \cite{zentsai2004} among others]. Similarly, the criterion $\mu_1$ corresponds to a weighted sum of $I$-optimality criteria in the case $\mathcal{X} = \mathcal{Z}$. 

 \smallskip
 
(c)  
%Criteria of the type \eqref{p_crit} and \eqref{infty_crit} have   been considered by \cite{dettobri1999} in the case of one regression model.  However, %these results cannot be extended for the criteria defined in Definition \ref{def_mu}, because  the 
It follows from Minkowski inequality  that in general the pair of the optimal designs 
 for the individual  models  $m_i \ (i=1, 2)$,  is not necessarily $\mu_p$-optimal in terms of Definition \ref{def_mu}.
\end{rem}

In some applications it might not be possible to conduct the experiments for both groups simultaneously. This situation arises, for example, in the analysis of clinical trials where data from different sources is available and one trial has already been conducted, while the other is planned in order to compare the corresponding two response curves. In this case only one design (for one group), say $\xi_1$, can be chosen, while the other is fixed, say $\eta$. The corresponding criteria are defined as
\be \label{nu1}
\nu_p (\xi_1) = \mu_p (\xi_1, \eta), \qquad p \in [1, \infty],
\ee
 and $\nu_p$ is minimized in the class of all designs on the design space $\mathcal{X}$ with non-singular information matrix $M_1(\xi_1, \vartheta_1)$. The corresponding  design minimizing $\nu_p$  is called $\nu_p$-optimal throughout this paper.

 \noindent \section{Optimal Design Theory}
\def\theequation{3.\arabic{equation}}
\setcounter{equation}{0}
\label{sec3}

A main tool of optimal design theory are equivalence theorems which, on the one hand, provide a characterization of the optimal design and, on the other hand, are the basis of many procedures for their numerical construction  [see for example \cite{detpep2008} or  \cite{yu2010},  \cite{yanbie2013}]. The following two results give the equivalence theorems for the $\mu_p$-criterion in the cases $p \in [1,\infty)$ (Theorem \ref{eqtheo_p}) and $p = \infty$ (Theorem \ref{eqtheo_inf}). Proofs can be found in Section \ref{sec6}. Throughout this paper $\supp(\xi)$ denotes the support of the design $\xi$ on $\mathcal{X}$.

\begin{satz}\label{eqtheo_p}
Let $p \in [1, \infty)$. The   design $\xi^{\star, p} = (\xi^{\star, p}_1, \xi^{\star, p}_2)$  is  $\mu_p$-optimal if and only if the inequality
\be \label{equi_p}
 \int_{\mathcal{Z}} \vp(t, \xi^{\star, p}_1, \xi^{\star, p}_2)^{p-1} \Bigl(  \frac{\gamma_1}{\sigma^2_1}{\vp}^2_1(t_1, t, \xi^{\star, p}_1) +  \frac{\gamma_2}{\sigma^2_2}{\vp}^2_2(t_2, t, \xi^{\star, p}_2) \Bigr) d\lambda(t)  - \mu^p_{p}(\xi^{\star, p}_1, \xi^{\star, p}_2) \leq 0
\ee
holds for all $t_1, t_2 \in \mathcal{X}$, where
\be \label{tilde_phi}
{\vp}_i(d, t, \xi^{\star, p}_i) = \frac{\sigma^2_i}{\gamma_i} f^T_i (d) M^{-1}_i ( \xi^{\star, p}_i, \vartheta_i) f_i(t),~i=1,2;
\ee
and the function $\vp(t, \xi^{\star, p}_1, \xi^{\star, p}_2)$  is defined in \eqref{phi_def}. Moreover, equality is achieved in \eqref{equi_p} for any $(t_1, t_2) \in \supp(\xi^{\star, p}_1) \times \supp(\xi^{\star, p}_2)$.

\end{satz}

\begin{satz}\label{eqtheo_inf}
The design  $\xi^{\star, \infty} = (\xi^{\star, \infty}_1, \xi^{\star, \infty}_2)$  is  $\mu_{\infty}$-optimal
 if and only if there exists a measure $\varrho^{\star}$ on the set of the extremal points 
$$\mathcal{Z}(\xi^{\star, \infty} )= \left\{ t_0 \in \mathcal{Z} : \vp(t_0, \xi^{\star, \infty}_1,\xi^{\star,\infty}_2) =\sup_{t\in \mathcal{Z}}  \vp(t, \xi^{\star, \infty}_1, \xi^{\star,\infty}_2) \right\}$$
of the function $\vp(t, \xi^{\star, \infty}_1,\xi^{\star,\infty}_2)$, such that the inequality
\be \label{equi_inf}
\int_{\mathcal{Z} (\xi^{\star, \infty} ) }  \Bigl( \frac{\gamma_1}{\sigma^2_1}{\vp}^2_1(t_1, t, \xi^{\star, \infty}_1) +  \frac{\gamma_2}{\sigma^2_2}{\vp}^2_2(t_2, t, \xi^{\star, \infty}_2) \Bigr) d\varrho^{\star}(t)
- \mu_{\infty}(\xi^{\star, \infty}) \leq 0
\ee
holds for all $t_1, t_2 \in \mathcal{X}$, where the functions ${\vp}_1$ and  ${\vp}_2$  are defined in \eqref{tilde_phi}. Moreover,
equality is achieved in \eqref{equi_inf} for any $(t_1,t_2) \in \supp(\xi^{\star, \infty}_1) \times \supp(\xi^{\star, \infty}_2)$.

\end{satz}

Theorem \ref{eqtheo_p}   and  Theorem \ref{eqtheo_inf}  can be used to check the optimality of a given design. However, in general, the explicit calculation of locally $\mu_p$-optimal  designs is very difficult, and in order to investigate the quality of a (non-optimal) 
 designs  $\xi=(\xi_1, \xi_2)$ for the purpose of comparing curves,
  we consider its  $\mu_{p}$-efficiency  which is defined by
\be \label{eq:efficiency}
\mbox{eff}_{p}(\xi) = \frac{\mu_{p}(\xi^{\star, p})}{\mu_p(\xi)} \in [0, 1].
\ee
The following theorem provides a lower bound for the efficiency of a  design $\xi =(\xi_1, \xi_2)$ in terms of the functions appearing in the equivalence Theorems \ref{eqtheo_p} and \ref{eqtheo_inf}. It is remarkable that this bound does not require knowledge of the optimal design.

\begin{satz}\label{theo_eff}
Let $\xi=(\xi_1, \xi_2)$ be a pair  of designs with non singular information matrices $M_1(\xi_1, \vartheta_1) $, $M_2(\xi_2, \vartheta_2) $.
\begin{enumerate}
\item[(a)]  If $p\in [1, \infty)$, then
\be \label{eff_bound_p}
\small
  {\rm eff}_{p}(\xi)  \geq \frac{\mu^p_p(\xi)}{ \max_{t_1, t_2 \in \mathcal{X}} \int_{\mathcal{Z}} \vp(t, \xi_1, \xi_2)^{p-1} \left( \frac{\gamma_1}{\sigma^2_1}\left({\vp}_1(t, t_1, \xi_1) \right)^2 + \frac{\gamma_2}{\sigma^2_2}\left({\vp}_{2} (t, t_2, \xi_2) \right)^2 \right) d\la(t)}. 
\ee
\item[(b)]  If $p= \infty$, then
\be \label{eff_bound_inf}
\small
{\rm eff}_{\infty}(\xi) \geq \frac{\mu_{\infty}(\xi)}{\max_{t\in \mathcal{Z}(\xi)} \max_{t_1, t_2 \in \mathcal{X}} \frac{\gamma_1}{\sigma^2_1}{\vp}^2_1(t_1, t, \xi_1) +\frac{\gamma_2}{\sigma^2_2} {\vp}^2_2(t_2, t, \xi_2)}.
\ee
\end{enumerate}
\end{satz}

Now, we consider the case where one design $\eta$ is already fixed and the criterion can only be optimized by the other design. The proofs of the following two results are omitted since they are similar to the proofs of Theorems \ref{eqtheo_p} and \ref{eqtheo_inf}. 
\begin{satz}\label{eqtheo_one_p}
Let $p \in [1, \infty)$. The design $\xi_1^{\star, p}$ is  $\nu_p$-optimal  if and only if the inequality
\be \label{equi_p_one}
 \int_{\mathcal{Z}} \vp^{p-1} (t, \xi^{\star, p}_1, \eta)\Bigl( \frac{\gamma_1}{\sigma^2_1} {\vp}^2_1(t_1, t, \xi^{\star, p}_1) +\vp_2(t, t, \eta) \Bigr) d\lambda(t) 
  - \nu^p_{p}(\xi^{\star, p}_1) \leq 0
\ee
holds for all $t_1 \in \mathcal{X}$, where ${\vp}_i$  and $\vp$ are defined in \eqref{tilde_phi} and \eqref{phi_def}, respectively.
Moreover, equality is achieved in \eqref{equi_p_one} for any $t_1 \in \supp(\xi^{\star, p}_1) $.
\end{satz}

\begin{satz}\label{eqtheo_one_inf}
The design $\xi_1^{\star, \infty}$ is  $\nu_{\infty}$-optimal 
if and only if there exists a measure $\varrho^{\star}$ on the set of the extremal points 
$$\mathcal{Z}(\xi^{\star, \infty}_1 )= \left\{ t_0 \in \mathcal{Z} : \vp(t_0, \xi^{\star, \infty}_1,\eta ) =\sup_{t\in \mathcal{Z}}  \vp(t, \xi^{\star, \infty}_1, \eta) \right\}$$
of the function $\vp(t, \xi^{\star, \infty}_1,\eta)$, such that the inequality
\be \label{equi_inf_one}
\int_{\mathcal{Z} (\xi^{\star, \infty}_1 ) }  \frac{\gamma_1}{\sigma^2_1}{\vp}^2_1(t_1, t, \xi^{\star, \infty}_1) d\varrho^{\star}(t)
- \int_{\mathcal{Z} (\xi^{\star, \infty}_1 ) } \vp_1(t, t, \xi^{\star, \infty}_1)  d\varrho^{\star}(t) \leq 0
\ee
holds for all $t_1\in \mathcal{X}$, where the functions ${\vp}_1$ is defined in \eqref{tilde_phi}. Moreover,
equality is achieved in \eqref{equi_inf_one} for any $t_1 \in \supp(\xi^{\star, \infty}_1) $.
\end{satz}

 \noindent \section{Extrapolation}
\def\theequation{4.\arabic{equation}}
\setcounter{equation}{0}
\label{sec4}
In this section we consider the criterion $\mu_{\infty}$ and the case where the design space $\mathcal{X}$ and the space $\mathcal{Z}$  do not intersect, which
corresponds to the problem of comparing two curves for extrapolation. We are particularly interested in the difference between  curves modeled by 
the Michaelis Menten,
EMAX and loglinear model. It turns out that the results for these models can be easily obtained from a general result
 for weighted polynomial regression models, which is  of own interest and will be considered 
first. For this purpose assume that the design space $\mathcal{X}$ and the range ${\cal Z}$  are intervals, that is  $\mathcal{X}=[L_{\mathcal{X}}, U_{\mathcal{X}}]$, $\mathcal{Z}= [L_{\mathcal{Z}}, U_{\mathcal{Z}}]$ and that both regression models $m_1$ and $m_2$ are given by functions of 
the type
\be \label{wpoly}
m_i(t)= \omega_i(t) \sum_{j=0}^{p_i}  \vartheta_{ij}\,  t^j \quad i= 1, 2, 
\ee
where $\omega_1, \omega_2$ are positive weight functions on $\mathcal{X} \cup \mathcal{Z}$. The models $m_1, \, m_2$ are called weighted polynomial regression models
and in  the case of one model
several design problems have been discussed in the literature,  mainly for the $D$- and $E$-optimality criterion [see for example \cite{det1993}, \cite{hei1994}, \cite{ant2003}, 
\cite{chang2005b,chang2005a} or \cite{detttram2010}]. 
 It is easy to show that the systems $\{\omega_i(t) t^j | j= 0, \ldots, p_i\}$ are Chebyshev systems on the convex hull of  $\mathcal{X} \cup \mathcal{Z}  $, say
 conv$(\mathcal{X} \cup \mathcal{Z}) $, 
 which means that for any choice $ \vartheta_{i0}, \ldots ,  \vartheta_{ip_i}$ the equation 
 $\omega_i(t) \sum_{j=0}^{p_i}  \vartheta_{ij}  t^j  =0$   has at most $p_i$ solutions in conv$(\mathcal{X} \cup \mathcal{Z} ) $
  [see \cite{karstu1966}]. It then follows from this reference that there exist unique polynomials $\underline{v}_i(t)= \omega_i(t) \sum_{j=0}^{p_i} a_{ij}  t^j ,\, i = 1, 2$
satisfying the properties 
\begin{enumerate}
\item for all $t \in \mathcal{X}$ the inequality $| \underline{v}_i(t)| \leq 1$  holds.
\item there exist $p_i +1$ points $L_{\mathcal{X}} \leq t_{i0} < t_{i1} < \ldots < t_{ip_i} \leq U_{\mathcal{X}}$ such that $\underline{v}_i(t_{ij}) = (-1)^j$ for $j=0, \ldots, p_i$.
\end{enumerate}
The points $t_{i0}, \ldots, t_{ip_i}$ are called Chebyshev points while $\underline{v}_i$ is called Chebyshev or equioscillating polynomial. 
 The following results give an  explicit  solution of the $\mu_{\infty}$-optimal design problem  if the functions $m_1$ and $m_2$ are weighted polynomials.

\begin{satz} \label{theo_extra_wpoly}
Consider the weighted polynomials \eqref{wpoly} with differentiable, positive weight functions $\omega_1, \, \omega_2$ {\bf such that for $\omega_i(t) \neq c \in \mathbb{R}$ $\{1, \omega_i(t), \omega_i(t)t, \ldots, \omega_i(t)t^{2p_i-1}\}$ and \\
$\{1, \omega_i(t), \omega_i(t)t, \ldots, \omega_i(t)t^{2p_i}\}$ are Chebshev systems ($i=1,2$).} Assume that $\mathcal{X} \cap \mathcal{Z}=
[L_{\mathcal{X}}, U_{\mathcal{X}}] \cap  [L_{\mathcal{Z}}, U_{\mathcal{Z}}]= \emptyset$.
\begin{enumerate}
\item If $U_{\mathcal{X}} <  L_{\mathcal{Z}}$ and $ \omega_1, \omega_2$ are strictly increasing on $\mathcal{Z}$, the support points of the $\mu_{\infty}$- optimal   design $\xi^{\star, \infty}= (\xi^{\star, \infty}_1, \xi_2^{\star, \infty})$ are given by the extremal points of the Chebyshev polynomial $\underline{v}_1(t)$ for $\xi^{\star, \infty}_1$ 
and $\underline{v}_2(t)$ for $\xi^{\star, \infty}_2$ with corresponding weights
\be \label{eq:extra_wpoly_weig}
\xi_{ij} = \frac{| L_{ij} ({U}_{\mathcal{Z}})|}{ \sum_{k=0}^{p_i} | L_{ik}({U}_{\mathcal{Z}})|}\quad j= 0, \ldots, p_i,\quad  i= 1, 2.
\ee
Here  $L_{ij}(t)
= \omega_i(t)  \sum_{j=0}^{p_i} \ell_{ij} t^j $ is the $j$-th Lagrange interpolation polynomial with knots $t_{i0}, \ldots, t_{ip_i}$, $i=1, 2$ defined
by the properties  $L_{ij}(t_{ik})  = \delta_{jk} $, $j,k=1,\ldots , p_i$ (and $ \delta_{jk} $ denotes the Kronecker symbol).
\item If $L_{\mathcal{X}} > U_{\mathcal{Z}}$  and $ \omega_1, \omega_2$ are strictly decreasing on $\mathcal{Z}$, the support points of the $\mu_{\infty}$-optimal design $\xi^{\star, \infty}= (\xi^{\star, \infty}_1, \xi_2^{\star, \infty})$ are given by the extremal points of the Chebyshev polynomial  $\underline{v}_1(t)$ for $\xi^{\star, \infty}_1$ and $\underline{v}_2(t)$ for $\xi^{\star, \infty}_2$ with corresponding weights
\beo
\xi_{ij} = \frac{| L_{ij} ({L}_{\mathcal{Z}})|}{ \sum_{k=0}^{p_i} | L_{ik}({L}_{\mathcal{Z}})|},\quad j= 0, \ldots, p_i,\quad  i= 1, 2.
\eeo
\end{enumerate}
\end{satz}

\parskip 12pt

\begin{exam}\label{exam_extra_poly}\rm
If both regression models $m_1$ and $m_2$ are given by polynomials of degree $p_1$ and $p_2$, we have
 $\omega_1\equiv \omega_2\equiv1$ and the $\mu_{\infty}$-optimal design can be described even more  explicitly. 
For the sake of brevity we only consider the case $U_{\mathcal{X}}  < L_{\mathcal{Z}}$. 
According to Theorem \ref{theo_extra_wpoly}  $\xi_1^{\star,  \infty} $ and $\xi_2^{\star, \infty}$ are supported at the extremal points of the 
polynomials 
  $\underline{v}_1(t)$ and $\underline{v}_2(t)$.  If $\omega_1\equiv \omega_2\equiv1$  these are given by  the Chebyshev polynomials  of the first kind on the interval $[L_{\mathcal{X}},U_{\mathcal{X}}]$, that is 
$$\underline{v}_1(t)= T_{p_1}\left(\frac{2t- (U_{\mathcal{X}} + L_{\mathcal{X}})}{U_{\mathcal{X}} - L_{\mathcal{X}}}\right) \quad \mbox{and} \quad \underline{v}_2(t) = T_{p_2}\left(\frac{2t- (U_{\mathcal{X}} + L_{\mathcal{X}})}{U_{\mathcal{X}} - L_{\mathcal{X}}}\right), $$
where $T_p(x)= \cos(p\arccos x) $, $x \in [-1,1]$.
Consequently, the component  $\xi^{\star, \infty}_i$  of the optimal  design is supported at  the $p_i +1$ Chebyshev points 
$$t_{ij}=  \frac{(1 -\cos(\tfrac{j}{p_i}\,\pi)) U_{\mathcal{X}} + (1+\cos (\frac{j}{p_i}\,\pi)) L_{\mathcal{X}}}{2}, \quad j= 0, \ldots, p_i $$
with corresponding weights
\be \label{eq:extra_poly_weig}
 \xi_{ij} =  \frac{| L_{ij}({U}_{\mathcal{Z}})|}{ \sum_{k=0}^{p_i} | L_{ik}({U}_{\mathcal{Z}})|}, \quad j= 0, \ldots, p_i
 \ee
where 
\beo
L_{ij}(t) = \prod_{k=0, k\neq j }^{p_i}\frac{t- t_{ik}}{t_{ij} - t_{ik}}.
\eeo
is the Lagrange interpolation polynomial at the knots $t_{i0}, \ldots, t_{ip_i}$. 
\end{exam}

While  Theorem \ref{theo_extra_wpoly} and Example \ref{exam_extra_poly} are of own interest, they turn out to be particularly useful to find 
$\mu_\infty$-optimal  designs for some commonly used dose response models.
To be precise we consider the  Michaelis Menten model
\be\label{eq:mod_MM}
m(t, \vartheta) = \frac{\vartheta_1 t}{\vartheta_2 + t}
\ee 
the loglinear model with fixed  parameter $\vartheta_3$ 
\be \label{eq:mod_LogLin}
m (t, \vartheta)= \vartheta_1 + \vartheta_2 \log(t + \vartheta_3)
\ee
and the EMAX model 
\be \label{eq:mod_Emax}
m(t, \vartheta)= \vartheta_1 +  \frac{\vartheta_2 t}{\vartheta_3 + t}.
\ee
The following result specifies the $\mu_\infty$-optimal  designs for the comparison of curves if  $\mathcal{X} \cap \mathcal{Z} = \emptyset$ and $m_1$ and $m_2$ are given by any of these models.

\medskip

\begin{kor} \label{cor_extra_nonlin}
Assume that  the regression models $m_1$ and $m_2$  are given by one of the models  \eqref{eq:mod_MM} -  \eqref{eq:mod_Emax}, 
 $L_{\mathcal{X}}\geq 0$ and   $U_{\mathcal{X}} <  L_{\mathcal{Z}}$. 
The $\mu_{\infty}$-optimal  design is given by $\xi^{\star, \infty} = (\xi^{\star, \infty}_1, \xi^{\star, \infty}_2)$, where $\xi^{\star, \infty}_i $ 
is given  by 
\beo
\xi^{\star, \infty}_{MM} = \begin{pmatrix}  \tfrac{\vartheta_2 U_{\mathcal{X}} (\sqrt{2}-1) }{(2-\sqrt{2}) U_{\mathcal{X}} + \vartheta_2  } & U_{\mathcal{X}} \\
\tfrac{ \vartheta_2 (U_{\mathcal Z} - U_{\mathcal X})}{ U_{\mathcal X} U_{\mathcal Z} (3 \sqrt{2} - 4) + \vartheta_2 (\sqrt{2} U_{\mathcal Z} - (4- 2\sqrt{2}) U_{\mathcal X})}& \tfrac{ (\sqrt{2}-1) \left[ (2- \sqrt 2) U_{\mathcal X} U_{\mathcal Z}  + \vartheta_2 (U_{\mathcal Z} - (\sqrt 2 -1)U_{\mathcal X})\right]}{ U_{\mathcal X} U_{\mathcal Z} (3 \sqrt{2} - 4) + \vartheta_2 \left [\sqrt{2} U_{\mathcal Z} - (4- 2\sqrt{2}) U_{\mathcal X}\right]}
\end{pmatrix},
\eeo
if $m_i$ is  the Michaelis Menten model and $L_{\mathcal{X}} > 0$, by 
\beo
\xi^{\star, \infty}_{LogLin} = \begin{pmatrix} L_{\mathcal{X}} & U_{\mathcal{X}} \\
\frac{\exp(U_{\mathcal{Z}})- \exp(U_{\mathcal{X}})}{2\exp(U_{\mathcal{Z}})-(\exp(L_{\mathcal{X}})+\exp(U_{\mathcal{X}}))} & \frac{\exp(U_{\mathcal{Z}})- \exp(L_{\mathcal{X}})}{2\exp(U_{\mathcal{Z}})-(\exp(L_{\mathcal{X}})+\exp(U_{\mathcal{X}}))} \end{pmatrix},
\eeo
if $m_i$ is   the loglinear model and by  
\beo
\xi^{\star, \infty}_{Emax} = \begin{pmatrix}L_{\mathcal{X}} & \frac{2 U_{\mathcal{X}}  L_{\mathcal{X}} +  ( U_{\mathcal{X}}  +L_{\mathcal{X}})\vartheta_3}{2\vartheta_3 +U_{\mathcal{X}} +  L_{\mathcal{X}}}  & U_{\mathcal{X}} \\
\frac{  (g(U_{\mathcal{Z}},U_{\mathcal{X}})  +g(U_{\mathcal{Z}},L_{\mathcal{X}})) g(U_{\mathcal{Z}},U_{\mathcal{X}})     }{L} 
&\frac{4 g(U_{\mathcal{Z}},U_{\mathcal{X}})  g(U_{\mathcal{Z}},L_{\mathcal{X}})  }{L}  &\frac{(g(U_{\mathcal{Z}},U_{\mathcal{X}})  +g(U_{\mathcal{Z}},L_{\mathcal{X}})) g(U_{\mathcal{Z}},L_{\mathcal{X}})   }{L} 
\end{pmatrix}
\eeo
if $m_i$ is the EMAX model. Here the function $g$ is defined by   $g(a,b) = \tfrac{a}{a  + \vartheta_3} - \tfrac{b}{b  + \vartheta_3} $ and $L$ is a normalizing constant, that is 
$L=g^2 (U_{\mathcal{Z}},U_{\mathcal{X}})  + 6 g(U_{\mathcal{Z}},U_{\mathcal{X}})  g(U_{\mathcal{Z}},L_{\mathcal{X}})  +g^2(U_{\mathcal{Z}},L_{\mathcal{X}})$.
\end{kor}

 \noindent \section{Numerical results}
\def\theequation{5.\arabic{equation}}
\setcounter{equation}{0}
\label{sec5}
In most cases of practical interest the $\mu_p$-optimal  design have to be found numerically. In the case $p < \infty$ the optimality criteria are in fact differentiable and several procedures can be used for this purpose [see \cite{detpep2008}, \cite{yang2010} or \cite{yanbie2013}]. In particular the optimality of the numerically constructed designs can be easily checked using the equivalence Theorem \ref{eqtheo_p}.
For this reason we concentrate on the case $p = \infty$ which is also probably of most practical interest, because it directly refers to the maximum width of the confidence band.  The $\mu_{\infty}$-optimality criterion is not necessarily differentiable. As a consequence there appears  the unknown measure $\varrho^{\star}$ in Theorem \ref{eqtheo_inf}, which has to
be calculated simultaneously with the optimal design in order check its $\mu_{\infty}$-optimality.  For this purpose we adapt a 
 procedure introduced  by \cite{wong1993}.  
 To be   precise recall the definition of $\tilde{\vp}_i$ in \eqref{tilde_phi},
 and consider an arbitrary design $\xi=(\xi_1, \xi_2)$ and an arbitrary measure $\varrho$ defined on the set of the extremal points  $\mathcal{Z}(\xi)$, then the  following inequality holds
\beao
&&\max_{t_1, t_2 \in \mathcal{X}}  \int_{\mathcal{Z}(\xi)} (  \tfrac{\gamma_1}{\sigma^2_1} {\vp}^2_1(t_1, t, \xi_1) + \tfrac{\gamma_2}{\sigma^2_2} {\vp}^2_2(t_2, t, \xi_2) ) d\varrho (t) \\
&\geq& \int_{\mathcal{X}} \int_{\mathcal{Z}(\xi)}  \tfrac{\gamma_1}{\sigma^2_1}{\vp}^2_1(t_1, t, \xi_1) d\varrho(t) d \xi_1(t_1) +  \int_{\mathcal{X}} \int_{\mathcal{Z}(\xi)}  \tfrac{\gamma_2}{\sigma^2_2}{\vp}^2_2(t_2, t, \xi_2) d\varrho(t) d \xi_2(t_2) \\
&=&  \int_{\mathcal{Z}(\xi)} \vp(t, \xi_1, \xi_2) d\varrho(t) =  \mu_{\infty}(\xi).
\eeao
On the other hand it follows from the equivalence Theorem \ref{eqtheo_inf} that the opposite inequality  also holds 
for the $\mu_{\infty}$-optimal  design $\xi^{\star, \infty}=(\xi^{\star, \infty}_1, \xi^{\star, \infty}_2)$ and the corresponding measure $\varrho^{\star}$ on $\mathcal{Z}(\xi^{\star, \infty})$ 
[see inequality \eqref{equi_inf}].  Consequently, the measure 
$\varrho^{\star}$ is the measure on $\mathcal{Z}(\xi^{\star, \infty})$ which minimizes the function 
\bea
N_{\infty}(\varrho,\xi^{\star, \infty}) &=& \max_{t_1, t_2 \in \mathcal{X}}  \int_{\mathcal{Z}(\xi^{\star, \infty})} (  \tfrac{\gamma_1}{\sigma^2_1}{\vp}^2_1(t_1, t, \xi^{\star, \infty}_1) +  \tfrac{\gamma_2}{\sigma^2_2}{\vp}^2_2(t_2, t, \xi^{\star, \infty}_2) ) d\varrho (t) 
\label{nfct} \\ 
&=&\max_{t_1 \in \mathcal{X}}  \tfrac{\sigma^2}{\gamma_1} f^T_1(t_1)M_1^{-1}(\xi_1^{\star, \infty})M_1(\varrho) M_1^{-1}(\xi_1^{\star, \infty}) 
f_1(t_1) \nonumber \\ 
&+&  \max_{t_2 \in \mathcal{X}} \tfrac{\sigma^2}{\gamma_2} f^T_2(t_2)M_2^{-1}(\xi_2^{\star, \infty})M_2(\varrho) M_2^{-1}(\xi_2^{\star, \infty}) f_2(t_2). \nonumber 
\eea
The  $\mu_\infty$-optimal design
$\xi^{\star, \infty}=(\xi^{\star, \infty}_1, \xi^{\star, \infty}_2)$  and the corresponding measures $\varrho^{\star}$ for the equivalence theorems are now 
calculated numerically in three steps using  Particle Swarm Optimization (PSO)  [see for example \cite{Clerc2006}]: 
\begin{enumerate}
\item We calculate the  $\mu_{\infty}$-optimal design $\xi^{\star, \infty}=(\xi^{\star, \infty}_1, \xi^{\star, \infty}_2)$   using  PSO.
\item We calculate numerically the set of extremal points $\mathcal{Z}(\xi^{\star, \infty})=\{z_1, \ldots, z_k\}$ of the function 
$ \vp(t, \xi^{\star, \infty}_1, \xi^{\star, \infty}_2)$.
\item We calculate numerically the measure $\varrho^{\star}$ on $\mathcal{Z}(\xi^{\star, \infty})=\{z_1, \ldots, z_k\}$ which minimizes the function
$N_{\infty}(\varrho,\xi^{\star, \infty}) $ defined in \eqref{nfct}  
using PSO.
\end{enumerate}

The  calculations are terminated if the lower bound for the efficiency in Theorem \ref{theo_eff} exceeds  a given   threshold, say  $0.99$.
In the following discussion  we consider the exponential, loglinear and EMAX  model with their corresponding parameter specifications 
depicted in Table \ref{tab_example_nonlin}. 
\begin{table}[t]
%\scriptsize{
\begin{center}
\begin{tabular}{|l|c|c|}
\hline
model & $m(t, \vartheta)$ & parameters \\
\hline
EMAX & $\vartheta_1 + \frac{\vartheta_2 t}{t + \vartheta_3}$ & $(0.2, 0.7, 0.2)$ \\
exponential & $\vartheta_1 + \vartheta_2 \exp(t / \vartheta_3) $& $(0.183, 0.017, 0.28)$\\
loglinear & $\vartheta_1 + \vartheta_2 \log(t + \vartheta_3) $ & $(0.74, 0.33,  0.2)$ \\
\hline
\end{tabular}
\caption{\it Commonly used dose  response models with their parameter specifications [from \cite{bretz2005}].\label{tab_example_nonlin} }
\end{center}
%}
\end{table}
These models  have been  proposed by \cite{bretz2005}  as a selection of commonly used models to represent 
dose response relationships on the dose range $[0,1]$.  These  authors also proposed a design which allocates $20\%$ of the patients to the dose levels $0 ,\, 0.05,\,  0.2 ,\, 0.6 $ and $1$, 
and which will be called standard design in the following discussion.
We consider $\mu_{\infty}$-optimal designs for the three combinations of these models, where the design space $\mathcal{X} =\mathcal{Z}= [0, 1]$. The variances $\sigma^2_1$ and $\sigma^2_2$ are equal and given by $\sigma^2= 1.478^2$ as proposed in  \cite{bretz2005} and we assume $\gamma_1=\gamma_2= 0.5$.  
The resulting  $\mu_{\infty}$-optimal  designs  are displayed in Table \ref{tab_mu_inf_opt_designs}. In the diagonal blocks 
we have two identical designs reflecting the fact that in this case $m_1=m_2$. These  designs are actually  the $D$-optimal designs for the
corresponding common model, which follows by a straightforward application of the famous equivalence theorem for $D$- and $G$-optimal
designs  [see \cite{kiewol1960}]. \\
In the other cases the optimal designs are obtained from Table  \ref{tab_mu_inf_opt_designs} as follows. 
For example, the $\mu_{\infty}$-optimal   design for the combination of the EMAX ($m_1$) and the exponential model ($m_2$)  
can be obtained from the right upper block. The first component is the design for the exponential model, 
which allocates  $40.3 \%\, , 27.4\% \,, 32.3\%$ of the patients to the dose levels $0.00,\, 0.74 ,\, 1.00$. The second component is the design for the EMAX model  which allocates $32.0\% \, ,28.2 \%$, $39.8 \%$ of the patients to the dose levels $ 0.00 , \,0.15 ,  \,1.00 $. Note that the optimal designs for the particular model vary with respect to the different combinations of the models. For example, the weights of the optimal design for the EMAX and   exponential model differ from the weights of the optimal design for the EMAX and  loglinear model. 
\begin{table}[t]
\begin{center}
%\scriptsize{
\begin{tabular}{|l||ccc||ccc||ccc|}
\hline
 $m_1$/  $m_2$& \multicolumn{3}{c||}{EMAX} & \multicolumn{3}{c||}{loglinear} &  \multicolumn{3}{c|}{exponential} \\
 \hline
\multirow{4}{*}{EMAX}  & 0.00 & 0.14  & 1.00  		&  0.00 &  0.22 &1.00  	&		0.00 & 0.74  & 1.00 \\ 
					 & $33.\bar{3}$ & $33.\bar{3}$ & $33.\bar{3}$ &			 34.0 & 32.5  & 33.5  & 		40.3 & 27.4 & 32.3 \\
					\cline{2-10}
					& 0.00 & 0.14  & 1.00  			& 0.00&  0.15 & 1.00	&		0.00 & 0.15 &  1.00\\
					  & $33.\bar{3}$& $33.\bar{3}$& $33.\bar{3}$ 			& 33.4 & 32.7& 33.9	&		32.0 & 28.2 & 39.8  \\
\hline\hline
\multirow{4}{*}{loglinear}  &  &  & &  0.00& 0.23 & 1.00&		 	0.00 &0.74 & 1.00\\
						&  &  & & 			$33.\bar{3}$&$33.\bar{3}$& $33.\bar{3}$ &			39.2 & 26.8 & 34.0\\
						\cline{5-10}
						&  &  & &	0.00 & 0.23 & 1.00& 			 0.00& 0.24 & 1.00 \\ 
						&  &  & &	 $33.\bar{3}$  & $33.\bar{3}$ & $33.\bar{3}$&					33.5 & 27.8 & 38.7 \\
\hline \hline
\multirow{4}{*}{exponential }  &  &  & & &  &   & 		 	0.00 &  0.75 & 1.00 \\
					&  &  & & &  &  & 				$33.\bar{3}$ & $33.\bar{3}$ & $33.\bar{3}$ \\
						\cline{8-10}
						&  &  &  &  &  & &	 0.00 &  0.75 & 1.00  \\ 
						&  &  &  &  &  & &	$33.\bar{3}$ & $33.\bar{3}$ & $33.\bar{3}$  \\
\hline
\end{tabular}
%}
\caption{\it $\mu_\infty$-optimal  designs  for  different model combinations. 
Upper rows: support points. Lower rows: weights given in percent ($\%$). \label{tab_mu_inf_opt_designs}}
\end{center}
\end{table}
\begin{figure}[t]
\begin{center}
\includegraphics[width= 0.4\textwidth]{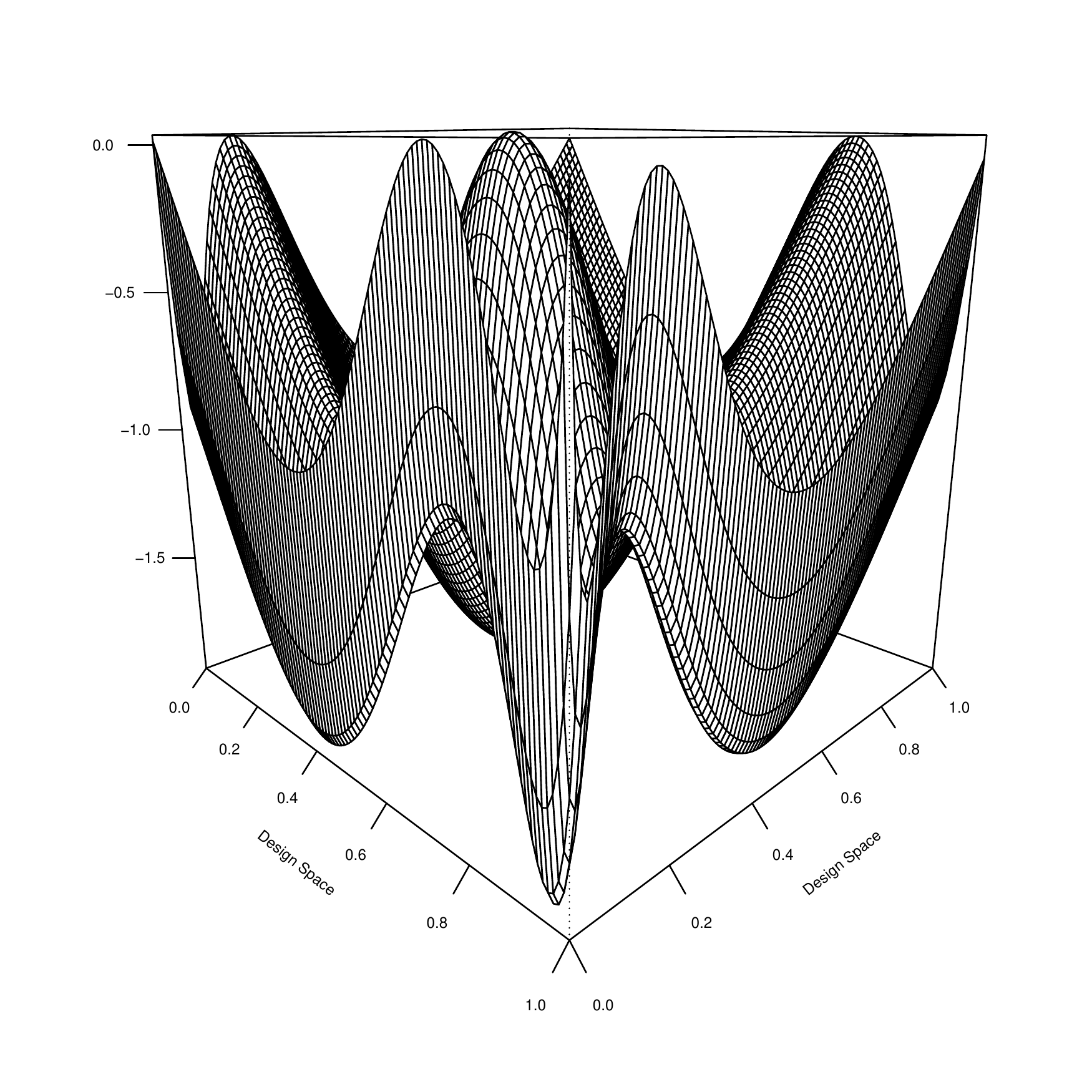} ~~~
\includegraphics[width= 0.4\textwidth]{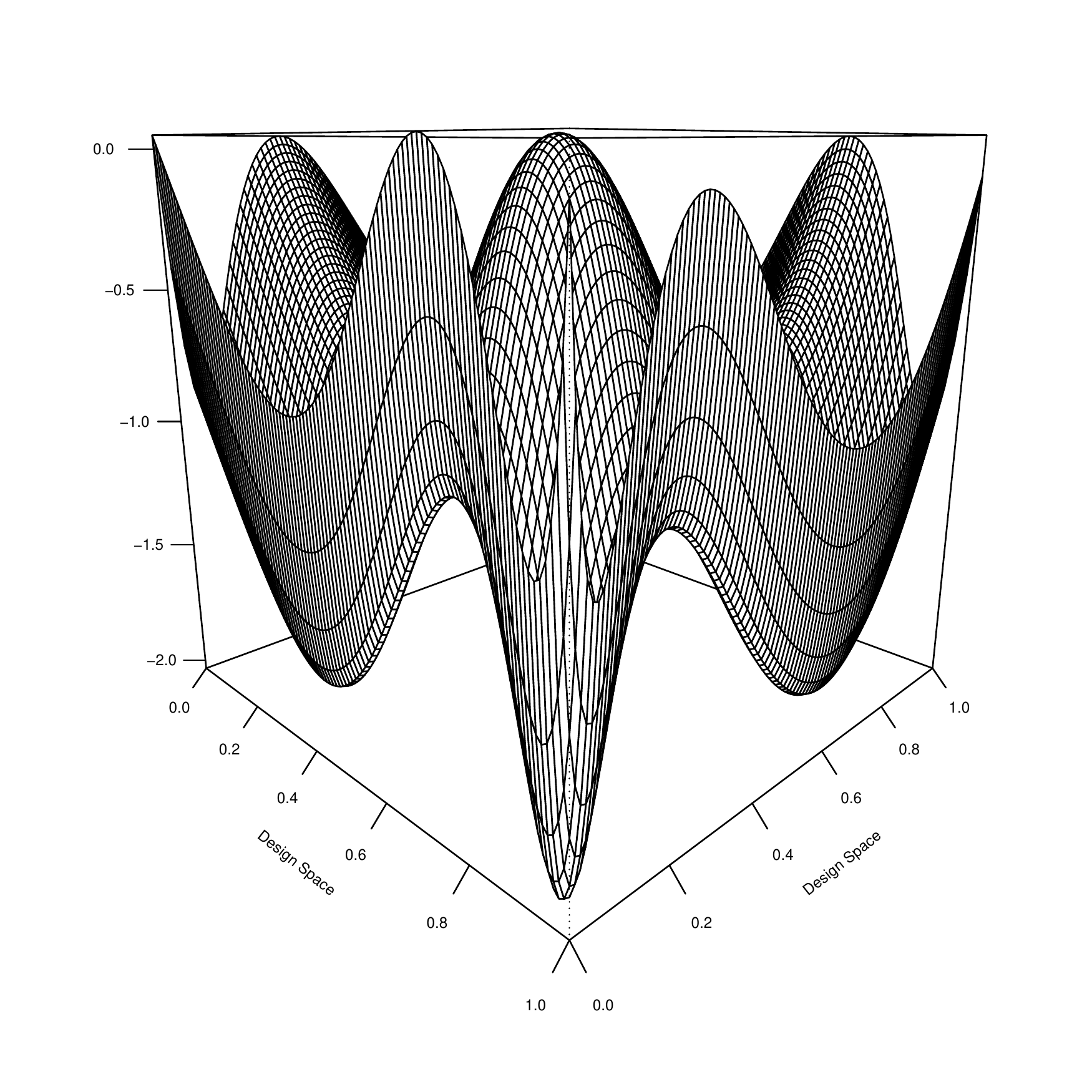} 
\caption{\it Illustration of Theorem \ref{eqtheo_inf}. The figures show the function on 
left hand side of inequality \eqref{equi_inf}. Left figure: The combination of exponential and EMAX model. Right figure: The combination of the loglinear and the exponential model. \label{fig:aeq_plots}}
\end{center}
\end{figure}
\begin{figure}[t]
\begin{center}
\includegraphics[width= 0.4\textwidth]{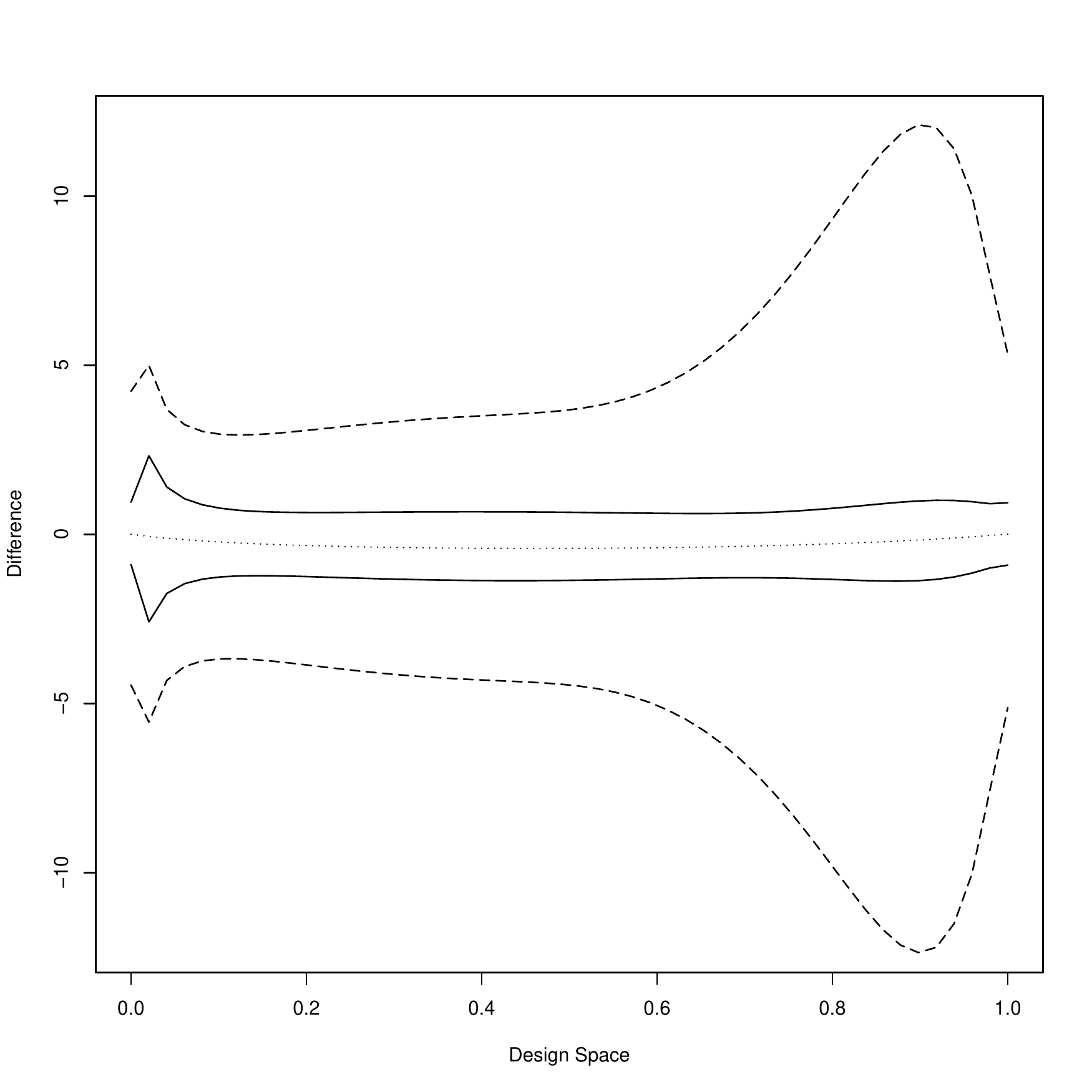}  ~~~~
\includegraphics[width= 0.4\textwidth]{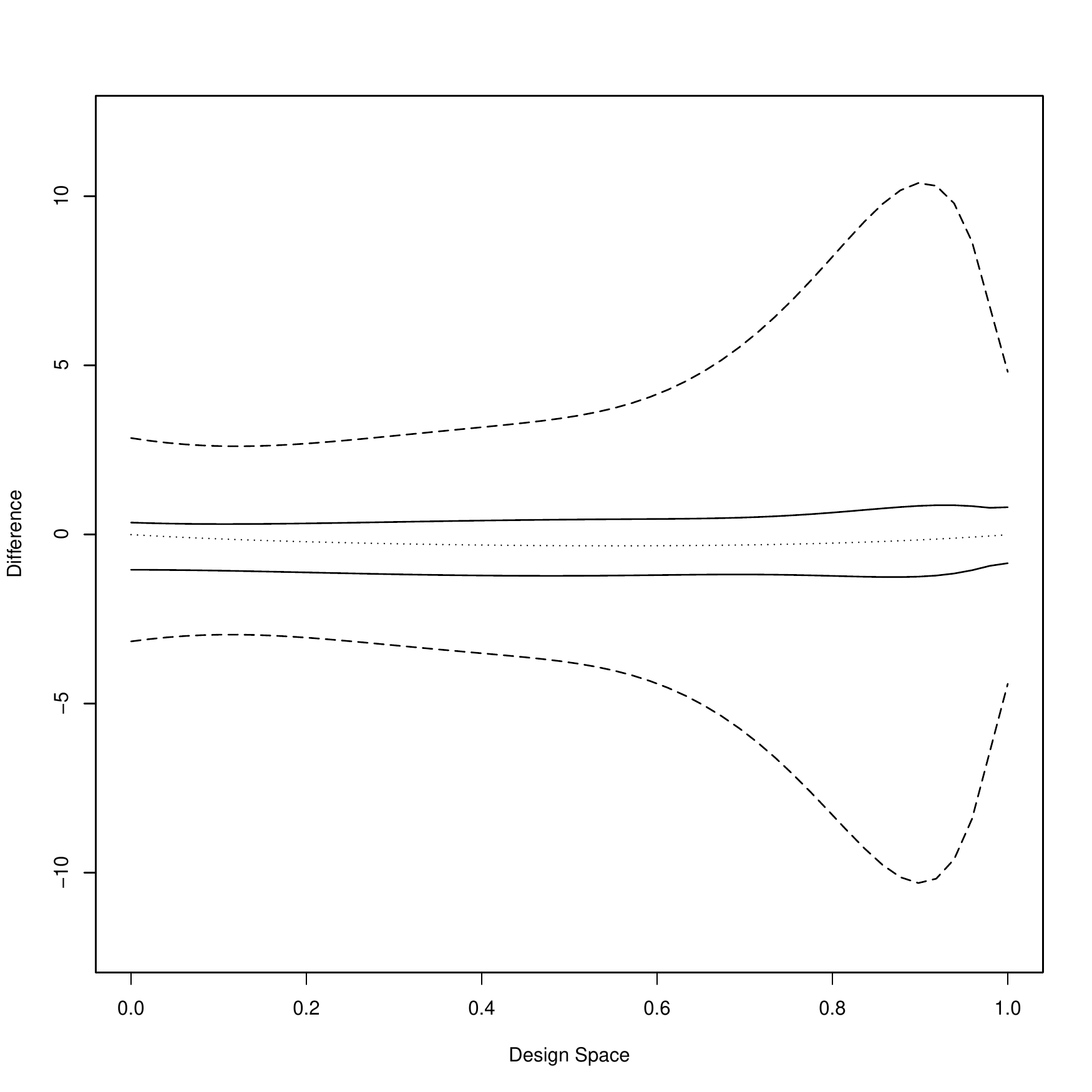} 
\caption{\it Confidence bands obtained from the $\mu_\infty$-optimal design (solid lines) and a standard design (dashed lines). 
The dotted line shows the true difference of the curves. Left figure: The combination of exponential and EMAX model. 
Right figure: The combination of the loglinear and the exponential model. \label{fig:kb_plots}}
\end{center}
\end{figure}
In Figure \ref{fig:aeq_plots} we demonstrate the application of the 
equivalence Theorem \ref{eqtheo_inf}  for the combinations EMAX and exponential model and exponential and loglinear model. 
Figure \ref{fig:kb_plots} presents the improvement of the confidence bands for the difference between the 
two regression functions if the $\mu_{\infty}$-optimal  design is
 used instead of a pair of the standard designs. The sample sizes in both groups are $n_1=100$ and $n_2=100$, respectively. The presented confidence bands are the averages of uniform confidence bands calculated by $100$ simulation runs. We observe that inference on the 
 basis of an  $\mu_\infty$-optimal   design yields a substantial reduction in the (maximal) width of the confidence band. \\
 \begin{table}[t]
\begin{center}
%\scriptsize{
\begin{tabular}{|c|c|c|c|}
\hline
model 1 / model 2 &  loglin / exp & loglin  /  EMAX & exp  /  EMAX \\
\hline
standard design & 58.85 & 72.83 & 59.00 \\
\hline 
$D$-optimal designs for EMAX  & 2.21 & 93.81 & 2.24 \\
\hline
$D$-optimal designs for loglinear   & 7.31 & 92.44 & 7.40\\
\hline			
$D$-optimal designs for exponential  & 15.08 & 3.72 & 4.29\\
\hline
$\nu_{\infty}$-optimal design (model 1 fixed) & 95.72 & 99.94 & 96.70\\
\hline
$\nu_{\infty}$-optimal design (model 2 fixed) &96.63 &  99.96 & 96.00 \\
\hline
\end{tabular}
%}
\caption{ {\it The $\mu_\infty$-efficiencies (in $\%$) of the standard design, pairs of  $D$-optimal designs (displayed in the  diagonal blocks of Table \ref{tab_mu_inf_opt_designs}) and the $\nu_{\infty}$-optimal designs (see Table \ref{tab:nu_inf_opt_designs})} . \label{tab_efficiency}}
\end{center}
\end{table}
Besides the comparison of the different confidence bands produced by the $\mu_{\infty}$-optimal design and the standard design proposed in \cite{bretz2005} we are able to compare them using the efficiency defined by \eqref{eq:efficiency}. The resulting efficiencies are depicted in the first row of Table \ref{tab_efficiency}.  We observe a substantial loss of efficiency if the standard design is used instead 
of a $\mu_\infty $-optimal design.
As  described in the previous paragraph the $\mu_\infty$-optimal  design is a  pair of two identical $D$-optimal designs  if both models coincide. 
These designs are depicted in the diagonal blocks of Table
\ref{tab_mu_inf_opt_designs}. In the row $2$-$4$ of Table \ref{tab_efficiency} we show the corresponding efficiencies, if these designs are used for the comparison of curves. 
For example, the $\mu_\infty$-optimal design for two EMAX models has $\mu_\infty$-efficiencies  $2.21\%$, $93.81\%$ and $2.24\%$, if it 
is used for the comparison of the loglinear and exponential, the loglinear and EMAX
and  the exponential and EMAX  model, respectively. Note that the  pair of $D$-optimal designs for the EMAX or loglinear model
is more efficient than the standard designs, if these two models are under consideration. In all other cases these designs have very low efficiency and
cannot be recommended for the comparison of curves.
Finally, we consider the $\nu_{\infty}$-criterion defined in \eqref{nu1}  assuming that the design for one model is already fixed as
the $D$-optimal design and we calculate the corresponding  $\nu_{\infty}$-optimal designs, which are depicted in Table \ref{tab:nu_inf_opt_designs} 
for the six possible combinations. For example, the $\nu_{\infty}$-optimal design for the comparison  of the exponential and EMAX model 
where the   design for the exponential model is fixed as $D$-optimal design
puts weights $35.1\%$, $29.7 \%$  and $35.2 \%$ at
 the  points  $0.00$,  $0.14$ and $1.00$, respectively. The $\mu_\infty$-efficiencies of 
these designs are presented in the  row $5-6$ of  Table \ref{tab_efficiency} and we observe that these designs 
 have very good efficiencies for the comparison of curves.
\begin{table}[t]
\begin{center}
%\scriptsize{
\begin{tabular}{| l ||ccc||ccc||ccc|}
\hline
 $m_1$/  $m_2$ & \multicolumn{3}{c||}{EMAX} & \multicolumn{3}{c||}{loglinear}  & \multicolumn{3}{c|}{exponential}  \\
 \hline
\multirow{2}{*}{EMAX} &  0.00 & 0.14  & 1.00  &  0.00 & 0.15 & 1.00 	& 0.00 & 0.14 & 1.00 \\
					& $33.\bar{3}$ & $33.\bar{3}$ & $33.\bar{3}$ & 34.0 & 32.0 & 34.0& 35.1 & 29.7& 35.2 \\
\hline\hline
\multirow{2}{*}{loglinear} &  0.00 & 0.23 & 1.00   &  0.00& 0.23 & 1.00  & 0.00 & 0.22 &1.00 \\
						& 34.0 & 32.0 & 34.0  & $33.\bar{3}$ & $33.\bar{3}$ & $33.\bar{3}$& 36.0 & 28.0 & 36.0 \\
\hline\hline
\multirow{2}{*}{exponential}  & 0.00 & 0.76 &  1.00 & 0.00 &0.75& 1.00  &	 0.00 &  0.75 & 1.00  \\
 							& 35.7 & 28.6 & 35.7 & 36.7 &  26.6 & 36.7  & $33.\bar{3}$ & $33.\bar{3}$ & $33.\bar{3}$  \\
\hline
\end{tabular}
%}
\caption{{\it $\nu_{\infty}$-optimal designs, where  one design is given by the $D$-optimal design of the second model. The weights are given in percent ($\%$). }\label{tab:nu_inf_opt_designs}}
\end{center}
\end{table}

%loglinear &  & $\xi^{\star, \infty}_{\mbox{loglin}} =\begin{pmatrix}   0.000 & 0.223 &1.000 \\  35.987 & 28.024 & 35.989\end{pmatrix}$
% 			&	$\xi^{\star, \infty}_{\mbox{loglin}} =\begin{pmatrix}   0.000 & 0.230 & 1.000 \\  33.990 & 32.018 & 33.991\end{pmatrix}$\\
%\hline
%  exponential  &  $\xi^{\star, \infty}_{\mbox{exp}} =\begin{pmatrix}  0.0000 &0.7467& 1.000\\36.705 &  26.590 & 36.705\end{pmatrix}$ 
%		  & & $ \xi^{\star, \infty}_{\mbox{exp}} =\begin{pmatrix} 0.000 & 0.764 &  1.000 \\ 35.651 & 28.696 & 35.653  \end{pmatrix} $   \\
%\hline
%EMAX   &  $\xi^{\star, \infty}_{\mbox{emax}} =\begin{pmatrix}  0.000 & 0.151 & 1.000 \\ 34.007 & 31.987 & 34.007 \end{pmatrix}$ 
%		  & $ \xi^{\star, \infty}_{\mbox{emax}} =\begin{pmatrix} 0.000 & 0.1440 & 1.000  \\ 35.126 & 29.747 & 35.127  \end{pmatrix} $ &  \\
%\hline
\parskip 12 pt

\section{Optimal allocation to the two groups} \label{sec6a}
So far we have assumed that the sample sizes $n_1$ and $n_2$ in the two groups are fixed and cannot be chosen by the experimenter.
In this section we will briefly indicate some results, if optimization can also be performed with respect to the relative proportions $\gamma_1=n_1/(n_1+n_2)$
and  $\gamma_2=n_2/(n_1+n_2)$ for the two groups. Following the approximate design approach we  define $\gamma$ as a
probability measure with masses  $\gamma_1$ and $ \gamma_2 $ at the points  $0$ and $1$, respectively, and  a $\mu_{\infty}$-optimal design
as a triple  $\xi^{\star, \infty} = (\xi^{\star, \infty}_1, \xi^{\star, \infty}_2, \gamma^{\star})$, which minimizes the functional
\beao
\mu_\infty
(  \xi_1, \xi_2, \gamma)   
&=&  \sup_{t \in \mathcal{Z}}   \vp 
(t,\xi_1, \xi_2, \gamma) ,
\eeao
where 
\beao
\vp 
(t,\xi_1, \xi_2, \gamma) 
&=&    \frac{\sigma^2_1}{\gamma_1} f_1^T(t) M_1^{-1} (\xi_1,  \vartheta_1) f_1(t) +  \frac{\sigma^2_1}{\gamma_2} f_2^T(t) M_2^{-1} (\xi_2,  \vartheta_2) f_2(t).
\eeao
Similar arguments as given in the proof of Theorem \ref{eqtheo_p} give a characterization of the optimal designs.  The details are omitted for the sake of brevity.\\

\begin{satz}\label{eq_theo_lambda_flex_inf}
A  design  $\xi^{\star, \infty} = (\xi^{\star, \infty}_1, \xi^{\star, \infty}_2, \gamma^{\star})$ is $\mu_{\infty}$-optimal if and only if there exists a measure $\varrho^{\star}$ on  
the set
\beo
\mathcal{Z}( \xi^{\star, \infty}_1, \xi^{\star, \infty}_2, \gamma^{\star}) = \{t \in \mathcal{Z} :\mu_{\infty}( \xi^{\star, \infty}_1, \xi^{\star, \infty}_2, \gamma^{\star}) =   
\vp 
(t,\xi^{\star, \infty}_1, \xi^{\star, \infty}_2, \gamma^{\star})
  \}
\eeo
such that the inequality
\be
\int_{\mathcal{Z}(\xi^{\star, \infty}_1, \xi^{\star, \infty}_2, \gamma^{\star}) } \tfrac{I\{\omega=0\}}{{\sigma_1^2}}\vp_1^2(t, t_1, \xi_1^{\star, \infty})+  \tfrac{I\{\omega=1\}}{{\sigma_2^2}}\vp_2^2(t, t_2, \xi_2^{\star, \infty})  \,  d\varrho^{\star}(t) -
 \mu_{\infty} (\xi^{\star, \infty}_1, \xi^{\star, \infty}_2, \gamma^{\star} ) \leq 0 \label{aeneu1} \\
\ee
is satisfied for all $t_1, t_2 \in \mathcal{X}$ and $\omega \in \{0, 1\}$, where $\vp_i$ is defined in \eqref{tilde_phi} with $\gamma_i= \gamma^{\star}_i$. Moreover, equality is achieved in \eqref{equi_inf} for any $(t_1,t_2, \omega) \in \supp(\xi^{\star, \infty}_1) \times \supp(\xi^{\star, \infty}_2) \times \{0,1\}$.
 \end{satz}
 %
%\parskip 12 pt
%\begin{figure}[t]
%\begin{center}
%\includegraphics[height= 0.4\textheight]{flex_vs_fix_emax_exp5.pdf}
%\caption{\it Confidence bands obtained from the $\mu_\infty$-optimal design (solid lines) with optimal $\gamma$ and from the $\mu_\infty$-optimal %design (dashed lines) with fixed $\gamma_1= \gamma_2=0.5$. The dotted line shows the true difference of the curves.}
%\label{fig_conf_gamma_flex_vs_fix}
%\end{center}
%\end{figure}

\begin{exam}\label{exam_gamma}\rm
The  $\mu_{\infty}$-optimal  design $ (\xi^{\star, \infty}_1, \xi^{\star, \infty}_2, \gamma^{\star})$ can be determined numerically in a similar way as 
described in Section \ref{sec5},  and we briefly illustrate some results for the comparison 
 of the EMAX model with the exponential model, where the parameters are given in Table \ref{tab_example_nonlin}. The 
variances  are $\sigma^2_1= 1.478^2$  in the first group and  $\sigma_2^2= 5 \cdot 1.478^2$ in the second group
and the optimal designs (calculated by the PSO) are presented in Table \ref{tab_gamma_flex}. Note that the optimal design allocates 
only $30.2\%$ of the observations to the first group. A comparison 
of the optimal designs from  Table \ref{tab_gamma_flex} with the corresponding  optimal designs from  Table \ref{tab_mu_inf_opt_designs}
(calculated under the assumptions  $\sigma^2_1=\sigma^2_2$ and $\gamma_1= \gamma_2=0.5$) 
shows that the support points are very similar, but there appear differences in the weights. 
% A comparison of the confidence bands obtained by the two designs is presented in Figure \ref{fig_conf_gamma_flex_vs_fix}.
\begin{table}
\small
\center
\begin{tabular}{|c|c|c|}
\hline
 $\gamma^* $ & $\xi^{\star, \infty}_1$  & $\xi^{\star, \infty}_2$ \\
 \hline
  (30.2, 69.8) & $ \begin{matrix}  0.00  & 0.15 & 1.00 \\ 32.4&  24.9&  42.7 \end{matrix} $  &  $\begin{matrix}  0.00 & 0.75 &1.00\\ 36.9& 30.4 & 32.7 \end{matrix}$\\
 \hline
\end{tabular}
\caption{\label{tab_gamma_flex} \it The $\mu_\infty$-optimal design  $ (\xi^{\star, \infty}_1, \xi^{\star, \infty}_2, \gamma^{\star})$   for the 
comparison of the EMAX- and exponential model, where optimization is also performed  with respect
to the relative  sample sizes $\gamma=(\gamma_1,\gamma_2)$  for the two groups.  The weights are given in $\%$.}
\end{table}
\end{exam}
\bigskip

{\bf Acknowledgements}. The authors would like to thank Martina
Stein, who typed parts of this manuscript with considerable
technical expertise. We are also grateful to Kathrin M\"ollenhoff for interesting discussions on the subject of comparing curves and for
computational assistance. This project has received funding from the European Union's 7th Framework Programme for research, 
  technological development and demonstration under the IDEAL Grant Agreement no 602552.
The work has  also been supported in part 
% the Collaborative Research Center ``Statistical modeling of
% nonlinear dynamic processes'' (SFB 823, Teilprojekt C2) of the German Research Foundation (DFG) 
   by the National Institute Of General Medical Sciences of the National 
Institutes of Health under Award Number R01GM107639. The content is solely the responsibility of the authors and does not necessarily
 represent the official views of the National
Institutes of Health.

  \noindent \section{Proofs}
\def\theequation{7.\arabic{equation}}
\setcounter{equation}{0}
\label{sec6}

Let $\Xi$ denote the space of all approximate designs on $\mathcal{X}$ and define
for $\xi_1, \xi_2 \in \Xi$
\begin{equation}\label{h1}
M(\xi_1, \xi_2, \vartheta_1, \vartheta_2) = \begin{pmatrix}
						\frac{\gamma_1}{\sigma^2_1} M_1(\xi_1, \vartheta_1) & 0_{s_1 \times s_2} \\
						0_{s_2 \times s_1} & \frac{\gamma_2}{\sigma^2_2} M_2(\xi_2, \vartheta_2)
						\end{pmatrix}
\end{equation}
as the block diagonal matrix with   information matrices $ \frac{\gamma_1}{\sigma^2_1} M_1(\xi_1, \vartheta_1)$ and $ \frac{\gamma_2}{\sigma^2_2} M_2(\xi_2, \vartheta_2)$ in the diagonal.  The set
$$
\mathcal{M}^{(2)} = \left\{M(\xi_1, \xi_2  , \vartheta_1, \vartheta_2)  		: \xi_1, \xi_2 \in \Xi
				\right\}
$$
is obviously a convex subset of the the set $ \NND(s_1 + s_2)$ of all non-negative definite $(s_1 + s_2)\times (s_1 + s_2)$ matrices. Moreover, if $\delta_t$ denotes the Dirac measure at the point $ t \in {\cal X}$ it is easy to see
that $\mathcal{M}^{(2)} $ is the convex hull of the set
$$\mathcal{D}^{(2)}= \left\{M(\delta_{t_1}, \delta_{t_2}, \vartheta_1, \vartheta_2) 	: t_1, t_2 \in \mathcal{X}
					\right\} ,$$
and that for any
 $p \in [1, \infty]$ the function $\mu_p(\xi)= \mu_p((\xi_1, \xi_2))$ defined in \eqref{p_crit} and \eqref{infty_crit} is convex on the set $\Xi \times \Xi$.

\bigskip

  \textbf{Proof of Theorem \ref{eqtheo_p}} 
Note that the  function $\vp$ in \eqref{phi_def} can be written as
\beo
\vp(t, \xi_1, \xi_2) = f^T(t) M^{-1}(\xi_1, \xi_2, \vartheta_1, \vartheta_2) f(t),
\eeo
where $f^T(t) = (f^T_1(t) , f^T_2(t))$ and $M(\xi_1, \xi_2) \in \mathcal{M}^{(2)}$ is defined in \eqref{h1}.
Similarly, we introduce for a matrix $M \in \mathcal{M}^{(2)}$ the notation $\Phi(M,t) = f^T(t) M^{-1}   f (t) $ and we rewrite the function $\mu_{p}(\xi_1, \xi_2)$  as
\be \label{mu_tilde_p}
\tilde{\mu}_{p}(M) = \Bigl( \int_{\mathcal{Z}}  \left(\Phi(M, t) \right)^p d\la(t)\Bigr)^{1/p} = \Bigl(\int_{\mathcal{Z}}  \left( f^T(t) M^{-1}   f (t)  \right)^p d\la(t)\Bigr)^{1/p}.
\ee
Because of the convexity of $\mu_p$ the  design $\xi^{\star, p}=(\xi_1^{\star, p}, \xi_2^{\star, p})$ is $\mu_p$-optimal if and only if the derivative of $\tilde{\mu}_{p}(M)$ evaluated in $M_0= M(\xi_1^{\star, p}, \xi_2^{\star, p}, \vartheta_1, \vartheta_2)$ is non-negative  for all directions $E_0=  E -M_0$, where $E \in \mathcal{M}^{(2)}$, i.e.
$
\partial \tilde{\mu}_{p}(M_0, E_0) \geq 0.
$
Since $\mathcal{M}^{(2)}= \mbox{conv}(\mathcal{D}^{(2)})$ it is sufficient to verify this inequality for all $E \in \mathcal{D}^{(2)}$.   \\
Assuming that integration  and differentiation  are interchangeable, the derivative at $M_0= M(\xi_1, \xi_2,  \vartheta_1, \vartheta_2)$  in direction  $E_0=  M(\delta_{t_1} \delta_{t_2} , \vartheta_1, \vartheta_2) -M_0$   is given by
\bea
\nonumber
&& \partial \tilde{\mu}_{p}(M_0,E_0) = \tilde{\mu}_{p}(M_0)^{1-p}  \int_{\mathcal{Z}} \left( f^T(t) M_0^{-1}   f (t)  \right)^{p-1} \left( - f^T(t) M_0^{-1} E_0 M^{-1}   f (t)  \right)  d \la(t) \\
 && ~~~= \tilde{\mu}_{p}(M_0)^{1-p} \int_{\mathcal{Z}} \left( f^T(t) M_0^{-1}   f (t)  \right)^{p} d\la(t) \nonumber \\
 & & ~~~ \quad -  \tilde{\mu}_{p}(M_0)^{1-p}  \int_{\mathcal{Z}}\left( f^T(t) M^{-1}   f (t)  \right)^{p-1} \left( f^T(t) M_0^{-1} M(\delta_{t_1}, \delta_{t_2} , \vartheta_1, \vartheta_2) M_0^{-1}   f (t)  \right)  d \la(t)  \nonumber\\
 && ~~~=\tilde{\mu}_{p}(M_0) -  \tilde{\mu}_{p}(M_0)^{1-p} \nonumber \\
  &&~~~ \int_{\mathcal{Z}} \Phi(M_0, t)^{p-1} \Bigl( \tfrac{\sigma^2_1}{\gamma_1}  (f^T_1(t)M_1^{-1}(\xi_1, \vartheta_1) f_2(t_1))^2 + \tfrac{\sigma^2_2}{\gamma_2} (f^T_2(t)M_2^{-1}(\xi_2, \vartheta_2) f_2(t_2))^2 \bigr) d\la(t) \nonumber \\
&& ~~~= \mu_{p}(\xi_1, \xi_2) \Bigl[ 1  -  {\mu}_{p}(\xi_1, \xi_2)^{-p} \int_{\mathcal{Z}} \beta (t,t_1,t_2) 
 d\la(t) \Bigr],
~~~~~~~~ \label{direct}
\eea
where the function $\beta$ is given by 
\be \label{beta}
 \beta (t,t_1,t_2)  =  \vp(t, \xi_1, \xi_2)^{p-1} (  \tfrac{\gamma_1}{\sigma^2_1}({\vp}_1(t, t_1, \xi_1) )^2 +  \tfrac{\gamma_2}{\sigma^2_2}({\vp}_{2} (t, t_2, \xi_2) )^2 ) .
 \ee
Consequently, the  design $\xi^{\star, p} = (\xi^{\star, p}_1, \xi^{\star, p}_2)$ is $\mu_p$-optimal if and only if the inequality
\be \label{ineq1}
\int_{\mathcal{Z}}   \beta (t,t_1,t_2) d\la(t) - \left(\mu_{p}(\xi^{\star, p}_1, \xi^{\star, p}_2)\right)^p \leq 0
\ee
is satisfied for all $t_1, t_2 \in \mathcal{X}$, which proves the first part of the assertion. \\
 It remains to prove that equality holds for any point $(t_1,t_2) \in \supp(\xi^{\star, p}_1) \times \supp(\xi^{\star, p}_2)$. For this purpose we assume the opposite, i.e. there exists a point
 $(t_1, t_2) \in \supp(\xi^{\star, p}_1) \times \supp(\xi^{\star, p}_2)$, such that there is strict inequality in \eqref{ineq1}.
This gives
\beo
\int_{\mathcal{X}} \int_{\mathcal{X}}\int_{\mathcal{Z}}   \beta (t,t_1,t_2)
 d\la(t) d \xi^{\star, p}_1(t_1) d \xi^{\star, p}_2(t_2)  \\
< \left(\mu_{p}(\xi^{\star, p}_1, \xi^{\star, p}_2)\right)^p.
\eeo
On the other hand, we have
\beao
&&\int_{\mathcal{X}} \int_{\mathcal{X}} \int_{\mathcal{Z}}    \beta (t,t_1,t_2) d\la(t) d \xi^{\star, p}_1(t_1) d \xi^{\star, p}_2(t_2)  
%\\  &=& \int_{\mathcal{Z}} \vp(t, \xi^{\star, p}_1, \xi^{\star, p}_2)^{p-1} \bigl\{ \int_{\mathcal{Z}} \tfrac{\gamma_1}{\sigma^2_1} \left({\vp}_1(t, t_1, \xi^{\star, p}_1) \right)^2 d \xi^{\star, p}_1(t_1) +  \int_{\mathcal{X}}  \tfrac{\gamma_2}{\sigma^2_2} ({\vp}_{2} (t, t_2, \xi^{\star, p}_2) )^2 d \xi^{\star, p}_2(t_2)  \bigr\} d\la(t) \\ &=&
  =\int_{\mathcal{Z}} \vp(t, \xi^{\star, p}_1, \xi^{\star, p}_2)^{p} d\la(t) =  \left(\mu_{p}(\xi^{\star, p}_1, \xi^{\star, p}_2)\right)^p.
\eeao
This contradiction shows that equality in \eqref{ineq1} must hold whenever  $(t_1,t_2) \in \supp(\xi^{\star, p}_1) \times \supp(\xi^{\star, p}_2)$.

\bigskip

 \textbf{Proof of Theorem \ref{eqtheo_inf}} 
 By the discussion at the beginning of the proof of Theroem \ref{eqtheo_p} the minimization of  the function $\mu_{\infty}(\xi_1, \xi_2)$ is equivalent to the minimization of
 \be \label{mu_tilde_inf}
 \tilde{\mu}_{\infty}(M) = \sup_{t \in \mathcal{Z}}  \Phi(M, t) = \sup_{t \in \mathcal{Z}} f^T(t) M^{-1}   f (t)
 \ee
 for $M \in \mathcal{M}^{(2)}$. From Theorem 3.5 in \cite{pshe1971} the subgradient of  $\tilde{\mu}_{\infty}(M)$ evaluated at a matrix $M_0$ in direction $E$ is given by
\beo
\mbox{D} \tilde{\mu}_{\infty}(M_0, E) = \Bigl\{\int_{\mathcal{Z}(M_0)} \partial \Phi(M_0, E, t) d\varrho (t) : \varrho \mbox{ measure on }  \mathcal{Z}(M_0) \Bigr\},
\eeo
where the set $ \mathcal{Z}(M_0)$ is defined by
$
\mathcal{Z}(M_0) = \left\{ t \in \mathcal{Z} : \tilde{\mu}_{\infty} (M_0) = \Phi(M_0, t)\right\},
$
and the derivative of $\Phi(M_0, t)$ in direction $E$ is given by
$
\partial \Phi(M_0,E, t) = - f^T(t) M_0^{-1} E M_0^{-1} f(t).
$
Applying the results from page $59$  in \cite{pshe1971} it therefore follows that the 
design  $\xi^{\star, \infty} = (\xi^{\star, \infty}_1, \xi^{\star, \infty}_2)$ is $\mu_{\infty}$-optimal if and only if there exists a measure $\varrho^{\star}$ on  $\mathcal{Z}( M(\xi^{\star, \infty}_1, \xi^{\star, \infty}_2, \vartheta_1, \vartheta_2)) $ such that the inequality
\beo
\int_{\mathcal{Z}(M_0) } \partial \Phi(M_0, E_0, t) d\varrho^{\star} (t)= \int_{\mathcal{Z}(M_0) } \partial \Phi(M_0, E , t) d\varrho^{\star} (t) + \int_{\mathcal{Z}(M_0) } f^T(t) M_0 ^{-1}   f (t) d\varrho^{\star}(t) \geq 0
\eeo
holds for all $E_0  = E- M_0$, $E \in \mathcal{M}^{(2)}$. Since $\mathcal{M}^{(2)}= \mbox{conv}(\mathcal{D}^{(2)})$ it is sufficient to consider the directions $E_0= E- M_0$, where $E \in \mathcal{D}^{(2)}$. Thus, this inequality is fulfilled if and only if there exists a measure $\varrho^{\star}$ on $\mathcal{Z}(M_0) = \mathcal{Z}(\xi^{\star, \infty})$, such that the inequality
\be \label{ineq_inf_1}
 \begin{split}
 \int_{\mathcal{Z}(\xi^{\star, \infty})  }  & f^T(t)  M^{-1} (\xi^{\star, \infty}_1, \xi^{\star, \infty}_2, \vartheta_1, \vartheta_2) M(\delta_{t_1}, \delta_{t_2}, \vartheta_1, \vartheta_2) M^{-1} (\xi^{\star, \infty}_1, \xi^{\star, \infty}_2, \vartheta_1, \vartheta_2)    d\varrho^{\star}(t) \\
\leq  & \int_{\mathcal{Z}(\xi^{\star, \infty}) } f^T(t) M ^{-1}   (\xi^{\star, \infty}_1, \xi^{\star, \infty}_2, \vartheta_1, \vartheta_2)   f (t) \, d\varrho^{\star}(t)  =\mu_{\infty} (\xi^{\star, \infty}_1, \xi^{\star, \infty}_2) 
 \end{split}
\ee
is satisfied for all $M(\delta_{t_1}, \delta_{t_2}, \vartheta_1, \vartheta_2) \in \mathcal{D}^{(2)}$. Observing the definition of $ \vp_i$ in \eqref{tilde_phi}, the left-hand
  part of \eqref{ineq_inf_1} can be rewritten as
$
 \int_{\mathcal{Z}(\xi^{\star, \infty}) }  \tfrac{\gamma_1}{\sigma^2_1}{\vp}^2_1(t_1, t, \xi^{\star, \infty}_1) +  \tfrac{\gamma_2}{\sigma^2_2}{\vp}^2_2(t_2, t, \xi^{\star, \infty}_2) \, d\varrho^{\star}(t),
$
and the inequality \eqref{ineq_inf_1} reduces to \eqref{equi_inf}. The remaining statement regarding the equality at the support points follows by the same arguments  as in the proof of Theorem \ref{eqtheo_p} and the details are omitted for the sake of brevity.
\medskip

 \textbf{Proof of Theorem \ref{theo_eff}}
 For both cases consider the function $(\tilde{\mu}_p(M))^{-1}$ where $\tilde{\mu}_p$ has already been defined in \eqref{mu_tilde_p} and \eqref{mu_tilde_inf}.
 Note that for each $t\in \mathcal{Z} $ the function $ M\to ( f(t)^TM^{-1}  f(t))^{-1}$ is concave [see \cite{pukelsheim2006}, p. 77], and consequently the function 
 $$(\tilde{\mu}_\infty  (M))^{-1} = {1 \over  \max_{t \in \mathcal{Z}}  f(t)^TM^{-1}  f(t)}  = \min_{t\in\mathcal{Z}} ( f(t)^TM^{-1}  f(t))^{-1}$$
 is also conave. The concavity of $(\tilde{\mu}_p(M))^{-1}$ in the case $1 \le p < \infty $ follows by similar arguments. 
For $p \in [1, \infty]$ the directional derivative of $(\tilde{\mu}_p(M))^{-1}$  at the point  $M_0$ in direction $E_0= M-M_0$ is given by
$$
\partial{(\tilde{\mu}_p}(M_0, E_0))^{-1} = - (\tilde{\mu}_p(M_0))^{-2} \partial{\tilde{\mu}_p}(M_0, E_0).
$$ 
We now consider the case  $p \in [1, \infty)$, the remaining case $p=\infty$ is briefly indicated at the end of this proof.
Observing \eqref{direct} a lower bound of the directional derivative of ${\tilde{\mu}_p}$ at 
$M_0= M(\xi_1, \xi_2,  \vartheta_1, \vartheta_2)$  in direction  $E_0=  M(\delta_{t_1} \delta_{t_2} , \vartheta_1, \vartheta_2) -M_0$ 
 is given by 
\beao
\partial{\tilde{\mu}_p}(M_0, E_0) &\ge &  \tilde\mu_{p}(M_0) \Bigl[ 1 -  {  \max_{t_1, t_2}  \int_{\mathcal{Z}}  \beta (t,t_1,t_2) d\la(t)  \over {\tilde\mu}_{p}^{p}  (M_0)} \Bigr]
 \eeao
 where $ \beta (t,t_1,t_2)  $ is defined in \eqref{beta}. 
Consequently, we have 
 \be \label{oben}
\partial{(\tilde{\mu}_p}(M_0, E_0))^{-1} \leq \frac{1}{\tilde{\mu}_{p}(M_0) }  \Bigl[  {  \max_{t_1, t_2}  \int_{\mathcal{Z}}  \beta (t,t_1,t_2)  d\la(t)  \over {\tilde\mu}_{p}^{p}  (M_0) }- 1  \Bigr].
\ee
Now, we  consider the matrices $M_0= M(\xi_1^{\star, p}, \xi_2^{\star, p},  \vartheta_1, \vartheta_2)$ of  the $\mu_p$-optimal
design and $M= M(\xi_1, \xi_2,  \vartheta_1, \vartheta_2) $ of any design $\xi=(\xi_1,\xi_2)$ with nonsingular information matrices $M_1(\xi_1, \vartheta_1) $
and $M_2( \xi_2,  \vartheta_2) $ and define the function 
$g_p(\alpha) = \tilde{\mu}_p((1-\alpha) M_0 + \alpha M))^{-1}$,
which is concave because of the concavity of $(\tilde{\mu}_p(M))^{-1}$. This yields   
\beao
 \frac{1}{\tilde{\mu}_{p}(M)} -\frac{1}{\tilde{\mu}_{p}(M_0)} &=& g_p(1) -g_p(0) \leq {\partial g_p(\alpha) \over \partial \alpha} \Bigl|_{\alpha=0} = 
\partial{(\tilde{\mu}_p}(M_0, E_0))^{-1}% \\
% &\leq& \frac{1}{\tilde{\mu}_{p}(M_0) }  \Bigl[  {  \max_{t_1, t_2}  \int_{\mathcal{Z}}  \beta (t,t_1,t_2)  \over {\mu}_{p}^{p}  (\xi_1, \xi_2)d\la(t) }- 1  \Bigr].
\eeao
Consequently, we obtain from \eqref{oben} the inequality 
$$
\mbox{eff}_{p}(\xi)  =  \frac{ \tilde{\mu}_{p}(M)}{\tilde{\mu}_{p}(M_0)}  \geq \frac{ \tilde \mu^p_p(M)}{   \max_{t_1, t_2} \int_{\mathcal{Z}}  \beta (t,t_1,t_2) d\la(t) }  ,
$$
which proves the assertion of Theorem \ref{theo_eff} in the case $1 \le p < \infty$. 
For the proof in the case $p=\infty$ we use similar arguments and  Theorem 3.2 in \cite{pshe1971}, which provides the upper bound 
 \be \label{eq_upper_inf}
 \partial{(\tilde{\mu}_{\infty}(M_0, E_0) )^{-1}}\leq \frac{1}{\tilde{\mu}_{\infty}(M_0)} \Bigl\{ \max_{d \in \mathcal{Z}(M_0)} \max_{t_1, t_2 \in \mathcal{X}} (f^T(d) M^{-1}_0  f(t_1, t_2))^2 - 1 \Bigr \},
 \ee
where $f(t_1, t_2)$ is defined by $f^T(t_1, t_2) = (f_1^T(t_1), f^T_2(t_2))^T$. The details are omitted for the sake of brevity.

\textbf{Proof of Theorem \ref{theo_extra_wpoly}} 
For the sake of brevity we now restrict ourselves to the proof of the first part of Theorem \ref{theo_extra_wpoly}. The second part can be proved analogously. 
Let $U_{\mathcal{X}} <  L_{\mathcal{Z}}$ and recall the definition of the function $ \vp(t, \xi_1, \xi_2)$ defined in \eqref{phi_def}.
% Assume both $\vp_1(t, \xi_1) $ and $\vp_2(t, \xi_2)$ are increasing in $t\in \mathcal{Z}$ for all pairs of designs $(\xi_1, \xi_2) \in \Xi(\mathcal{X}) \times  \Xi(\mathcal{X})$. 
The function $\vp(t, \xi_1, \xi_2)$ is obviously increasing on $\mathcal{Z}$, if the functions 
\beo
 \vp_i(t,t, \xi_i)= {\sigma_i^2\over \gamma_i} f^T_i(t) M_i^{-1}(\xi_i) f_i(t) = 
 {\sigma_i^2\over \gamma_i}  \omega_i^2(t)  (1,  t,  \ldots ,  t^{p_i}  ) M^{-1}_i(\xi_i) ( 1 ,  t,  \ldots , t^{p_i} )^T
\eeo
are increasing on $\mathcal{Z}$ for $i=1, 2$. In this case we have 
\be\label{eq:mu_ex}
\max_{t \in \mathcal{Z}} \vp(t, \xi_1, \xi_2) = \vp(U_{\mathcal{Z}}, \xi_1, \xi_2) =  \vp_1(U_{\mathcal{Z}}, \xi_1) + \vp_2(U_{\mathcal{Z}}, \xi_2) .
\ee
Because of this structure the components of the  optimal design can be calculated separately for $\vp_1$ and $\vp_2$.
Since both $\{\omega_1(t), \omega_1(t) t, \ldots, \omega_1(t) t^{p_1}\} $ and $\{\omega_2(t), \omega_2(t) t, \ldots, \omega_2(t) t^{p_2}\}$ are Chebyshev systems on $\mathcal{X} \cup \mathcal{Z}$ it follows from Theorem X.7.7 in \cite{karstu1966} that the support points of the  design
$\xi_i$ minimizing $\vp_i(U_{\mathcal{Z}}, \xi_i)$
 are given by the extremal points of the equioscillating polynomials $\underline{v}_i(t)$, while  the corresponding weights are given by \eqref{eq:extra_wpoly_weig}. \\
In order to prove the monotonicity of $\vp_i$, ($i=1,2$) let  $\xi_i$ denote a design with $k_i$ support points $t_{i0},  \ldots  , t_{i{k_i-1}} \in \mathcal{X}$ and corresponding weights $ \xi_{i0},  \ldots ,\xi_{i{k_i-1}}$. \\
% hier eingefügte Begründung
{\bf Since$\{1, \omega_i(t), \omega_i(t)t, \ldots, \omega_i(t)t^{2p_i-1}\}$ and $\{1, \omega_i(t), \omega_i(t)t, \ldots, \omega_i(t)t^{2p_i}\}$ are Chebshev systems  for $\omega_i(t) \neq c \in \mathbb{R}$, the complete class theorem of \cite{detmel2011} can be applied and it is sufficient to consider minimal supported designs $\xi_i$. Consequently, we set $k_i=p_i+1$.} 

Define $X_i = \left( \omega_i(t_{ik})  t^l_{ik} \right)_{k, l = 0, \ldots, p_i}$, then it is easy to see that the $j$th Langrange interpolation polynomial is given by $L_{ij}(t) =  e_j^T X_i^{-1} \left( \omega_i(t) ,\omega_i(t)  t,  \ldots , \omega_i(t)  t^{p_i} \right)^T$, where $e_j$ denotes the $j$th unit vector
(just check the defining condition $L_{ij}(t_{i\ell}) =\delta_{j\ell}$). With these notations the function  $\vp_i(t, \xi_i)$ can be rewritten as 
\be \label{eq_poly_lagr}
\begin{split}
\vp_i(t,t, \xi_i) &=  {\sigma_i^2\over \gamma_i}  \left( \omega_i(t) ,\omega_i(t)  t,  \ldots , \omega_i(t)  t^{p_i} \right)  X_i^{-T} W^{-1}_i X_i^{-1} \left( \omega_i(t) ,\omega_i(t)  t,  \ldots , \omega_i(t)  t^{p_i} \right)^T \\
&:= {\sigma_i^2\over \gamma_i}   \sum_{j=0}^{p_i}  \frac{1}{\xi_{ij}  } (L_{ij}(t) )^2,
\end{split}
\ee 
where $W_i= \mbox{diag}(\xi_{i0},  \ldots ,\xi_{ip_i})$. Now Cramer's rule  and a straightforward calculation yields the 
following representation for the Lagrange interpolation polynomial
{\footnotesize{
\beao
L_{ij}(t) &=& (-1)^{p_i - j}  \omega_i(t)\frac{\prod_{k=0, k\neq j }^{p_i} \omega_i (t_{ik})}{\prod_{k=0 }^{p_i} \omega_i (t_{ik})} \frac{ \det \begin{pmatrix} 1 & \ldots &  1 &1 & \ldots &1&   1  \\
							t_{i0} & \ldots&  t_{ij-1} & t_{ij+1} & \ldots & t_{ip_i}   & t \\
							\vdots & \ldots & \vdots & \vdots & \ldots  & \vdots & \vdots \\
							 t^{p_i}_{i0} & \ldots & t^{p_i}_{ij-1} & t^{p_i}_{ij+1} & \ldots  &t^{p_i}_{ip_i}   & t^{p_i}
							  \end{pmatrix} }{\det \begin{pmatrix} 1 & \ldots &  1 \\
							t_{i0} & \ldots &  t_{ip_i}  \\
							\vdots & \ldots & \vdots \\
							 t^{p_i}_{i0} & \ldots  & t^{p_i}_{ip_i} 	\end{pmatrix}} \\
							 &=& \frac{\omega_i(t)}{\omega_i(t_j)} \prod_{k=0, k\neq j }^{p_i}\frac{t- t_{ik}}{t_{ij} - t_{ik}}.
\eeao}}
Therefore the  partial derivative of $\vp_i(t, \xi_i)$ with respect to $t$ is given by 
 \beo
\frac{\partial}{\partial t} \vp_i(t,t, \xi_i) =  {\sigma_i^2\over \gamma_i}  \sum_{j=0}^{p_i}  \frac{2}{\xi_{ij}} (L_{ij}(t))^2 \Bigl(\frac{\omega'_i(t)}{\omega_i(t)} + \sum_{l=0}^{p_i} \frac{1}{t-t_{il}}\Bigr) .
\eeo
Note that $t_{il} < t$ for all $t_{il} \in \mathcal{X}$ and $t \in \mathcal{Z}$ and that both $\omega_i(t)$ and $\omega'_i(t)$ are positive. Consequently, the partial derivative is positive and the function $\vp_i(t, \xi_i)$ is increasing in $t \in \mathcal{Z}$.  Thus, the maximum value of $\vp_i(t, \xi_i)$ is attained in $U_{\mathcal{Z}} \in \mathcal{Z}$ and \eqref{eq:mu_ex} follows.

\parskip 12pt

\textbf{Proof of Corollary \ref{cor_extra_nonlin}}
For the sake of brevity we only prove the result for the EMAX model \eqref{eq:mod_Emax},  where it essentially follows by an application of Theorem \ref{theo_extra_wpoly} with $\omega(t)\equiv 1$. The proofs for the Michaelis Menten model and for the loglinear model are similar. In the 
the EMAX model the  gradient is given by 
$f(t, \vartheta) = (1, \frac{t}{t+ \vartheta_3} , \frac{-\vartheta_2 t}{(t+ \vartheta_3)^2} )$. 
Using the strictly increasing transformation
$z = v(t) = \frac{t}{\vartheta_3 + t}$
the function $f$ can be rewritten by
 $$
 f(t, \vartheta)  =  \begin{pmatrix} 1 & 0 &0 \\ 0 & 1 & 0 \\ 0 & -\frac{\vartheta_2}{\vartheta_3} &\frac{\vartheta_2}{\vartheta_3} \end{pmatrix} \begin{pmatrix}1 \\ z \\ z^2\end{pmatrix} := P_{\vartheta} \begin{pmatrix}1 \\ z \\ z^2\end{pmatrix}.
 $$
Thus, for an arbitrary design $\xi$ the function $f^T(t) M^{-1}(\xi) f(t)$ reduces to 
\beao
\vp(t, \xi) =  f^T(t) M^{-1}(\xi) f(t)& = &(1, z , z^2) P_{\vartheta}^T\left( P_{\vartheta} \tilde{M}(\tilde{\xi}) P^T_{\vartheta}\right)^{-1} P_{\vartheta} \, (1, z , z^2)^T \\
 &=& (1, z , z^2) \tilde{M}^{-1}(\tilde{\xi}) (1, z , z^2)^T = \tilde{\vp}(z, \tilde{\xi})
\eeao
where $\tilde{M}(\tilde{\xi}) = ( \int_{\mathcal{X}} z^{i+j} d\tilde{\xi}(z) )_{i, j= 0, 1,2}$ and $\tilde{\xi}$ is 
the design on the design space $\tilde{\mathcal{X}}= [ \frac{L_{\mathcal{X}}}{\vartheta_3 + L_{\mathcal{X}}},  \frac{U_{\mathcal{X}}}{\vartheta_3 +  U_{\mathcal{X}}}]$
induced from the actual design $\xi$ by the transformation  $z  = \frac{t}{\vartheta_3 + t}$. 
The function $\tilde{\vp}(z, \tilde{\xi})$ coincides with the variance function of a polynomial regression model with degree $2$ and constant weight function $\omega(t)   \equiv 1$. The corresponding design and extrapolation space are  given by $\tilde{\mathcal{X}}= [ \frac{L_{\mathcal{X}}}{\vartheta_3 + L_{\mathcal{X}}},  \frac{U_{\mathcal{X}}}{\vartheta_3 +  U_{\mathcal{X}}}]$ and  $\tilde{\mathcal{Z}}= [ \frac{ L_{\mathcal{Z}}}{\vartheta_3 + L_{\mathcal{Z}}},  \frac{ U_{\mathcal{Z}}}{\vartheta_3 +  U_{\mathcal{Z}}}]$, respectively. 
According to Example \ref{exam_extra_poly}  $(p_1=2)$ the component $\tilde \xi_i $ of the $\mu_\infty$-optimal
design is supported at the extremal points of the Chebyshev 
polynomial of the first kind on the interval $\mathcal{X}$, which are given by 
$$
\frac{L_{\mathcal{X}}}{\vartheta_3 + L_{\mathcal{X}}}, ~{1\over 2} 
\Bigl( \frac{L_{\mathcal{X}}}{\vartheta_3 + L_{\mathcal{X}}} +   \frac{U_{\mathcal{X}}}{\vartheta_3 +  U_{\mathcal{X}}} \Bigr),~
 \frac{U_{\mathcal{X}}}{\vartheta_3 +  U_{\mathcal{X}}}
$$
For the weights we obtain  by the same result 
$\xi_0= \tfrac{|L_0| }{L} , \xi_1= \tfrac{|L_1| }{L} , \xi_2= \tfrac{|L_2| }{L} $
where
\beao
|L_0| &=& \left( 2 \tfrac{ U_{\mathcal{Z}}}{ U_{\mathcal{Z}} + \vartheta_3} - \left(\tfrac{ U_{\mathcal{X}}}{ U_{\mathcal{X}} + \vartheta_3}  + \tfrac{ L_{\mathcal{X}}}{ L_{\mathcal{X}} + \vartheta_3}   \right) \right) \left(  \tfrac{ U_{\mathcal{Z}}}{ U_{\mathcal{Z}} + \vartheta_3}-\tfrac{ U_{\mathcal{X}}}{ U_{\mathcal{X}} + \vartheta_3} \right) \\
|L_1|&=&  4 \left(\tfrac{ U_{\mathcal{Z}}}{ U_{\mathcal{Z}} + \vartheta_3} -\tfrac{ L_{\mathcal{X}}}{ L_{\mathcal{X}} + \vartheta_3}  \right) \left(  \tfrac{ U_{\mathcal{Z}}}{ U_{\mathcal{Z}} + \vartheta_3}-\tfrac{ U_{\mathcal{X}}}{ U_{\mathcal{X}} + \vartheta_3} \right) \\
|L_2| &=& \left(  \tfrac{ U_{\mathcal{Z}}}{ U_{\mathcal{Z}} + \vartheta_3}-\tfrac{ L_{\mathcal{X}}}{ L_{\mathcal{X}} + \vartheta_3} \right) \left( 2 \tfrac{ U_{\mathcal{Z}}}{ U_{\mathcal{Z}} + \vartheta_3} - \left(\tfrac{ U_{\mathcal{X}}}{ U_{\mathcal{X}} + \vartheta_3}  + \tfrac{ L_{\mathcal{X}}}{ L_{\mathcal{X}} + \vartheta_3}   \right) \right)  \\
L&=& |L_0| +|L_1| +|L_2| .
\eeao
The support points of the of the $\mu_\infty$-optimal design $\xi$ are now obtained by the inverse of the transformation and the
assertion for the EMAX model follows from the definition of the function $g$ and a straightforward calculation.

\setstretch{1.25}
\setlength{\bibsep}{1pt}
\begin{small}
\bibliographystyle{apalike}
\bibliography{detschorn14}
\end{small}
\end{document}